\documentclass[journal]{IEEEtran}
\usepackage{marginnote}
\usepackage{cite}
\usepackage{balance} 
\usepackage{amsmath,amsfonts}
\usepackage{algorithmic}
\usepackage{array}
\usepackage[caption=false,font=normalsize,labelfont=sf,textfont=sf]{subfig}
\usepackage{textcomp}
\usepackage{stfloats}
\usepackage{url}
\usepackage{verbatim}
\usepackage{graphicx}
\usepackage{subcaption}
\usepackage{balance}
\usepackage{float}
\usepackage{amsthm}
\usepackage{microtype}
\usepackage{graphicx}
\usepackage{booktabs}
\usepackage{algorithm}
\usepackage{algorithmic}
\usepackage[hidelinks]{hyperref}
\usepackage{url}
\usepackage{dsfont} 
\usepackage{siunitx}
\usepackage{amsmath,amssymb,amsfonts}
\usepackage{cleveref}
\usepackage{mathrsfs}
\usepackage{hyperref}

\newtheorem{theorem}{Theorem}
\theoremstyle{remark}
\newtheorem{remark}{Remark}

\usepackage{bm}
\usepackage{amsmath}
\usepackage{algorithm}
\usepackage{algorithmic}

\IEEEoverridecommandlockouts

\begin{document}



\bstctlcite{IEEEexample:BSTcontrol}

\title{Optimizing the Driving Profile for Vehicle Mass Estimation - extended version}

\author{Le Wang, Jessica Ye, Michael Refors, Oscar Flärdh
and Håkan Hjalmarsson,~\IEEEmembership{Fellow,~IEEE}

\thanks{This work was supported by VINNOVA Competence Center AdBIOPRO (Contract 2016-05181), by the Swedish Research Council (Contract 2019-04956).}
\thanks{Le Wang and Håkan Hjalmarsson are with Division of Decision and Control Systems, School of Electrical Engineering and Computer Science, KTH Royal Institute of Technology, 10044 Stockholm, Sweden. Le Wang is also with Shanghai Jiao Tong University, 200240 Shanghai, China. Håkan Hjalmarsson is also with Centre for Advanced Bio Production, 10044 Stockholm, Sweden (email: \{le6, hjalmars\}@kth.se). Jessica Ye is a Master Thesis student at KTH (e-mail: jessicay@kth.se), Michael Refors and Oscar Flärdh are with Traton R\&D, 151 87 Södertälje, Sweden (e-mail: \{michael.refors, oscar.flardh\}@scania.com).}

}

\markboth{IEEE TRANSACTIONS ON CONTROL SYSTEMS TECHNOLOGY,~Vol.~1, No.~1, October~2025}%
{Shell \MakeLowercase{\textit{et al.}}: Bare Demo of IEEEtran.cls for IEEE Journals}

\maketitle

\begin{abstract}

Accurate mass estimation is essential for the safe and efficient operation of heavy-duty vehicles. While prior work has recognized the importance of acceleration profiles for estimation accuracy, the systematic design of driving profiles during transport has not been thoroughly investigated. This paper presents a framework for designing driving profiles to support accurate mass estimation. Based on application-oriented input design, it aims to meet a user-defined accuracy constraint under three optimization objectives: minimum time, minimum distance, and maximum accuracy (within a fixed time). It allows time- and distance-dependent bounds on acceleration and velocity, and incorporates actuator dynamics. The optimal profiles are obtained by solving concave optimization problems using a branch-and-bound method, with alternative rank-constrained and semi-definite relaxations also discussed. Theoretical analysis provides insights into the optimal profiles, including feasibility conditions, key ratios between velocity and acceleration bounds, and trade-offs between time- and distance-optimal solutions. The framework is validated through simulations and real-world experiments. Results show that the designed profiles are feasible and effective, enabling accurate mass estimation as part of normal transportation operations without requiring dedicated calibration runs. An additional contribution is a non-causal Wiener filter, with parameters estimated via the Empirical Bayes method, used to filter the accelerometer signal with no phase-lag. The method is developed for least-squares estimation but it is shown using the truck data that driving profile design also results in significant improvements for the accuracy of the recursive total least-squares method. Using the truck data we also show that least-squares estimation combined with Wiener filtering of the acceleration signal is competitive in comparison with this method.


\end{abstract}


\IEEEpeerreviewmaketitle


\section{Introduction}

\IEEEPARstart{A}CCURATE mass estimation is fundamental to the safe, efficient, and intelligent control of autonomous heavy-duty vehicles \cite{li2017two, mcintyre2009two}. This need is particularly pronounced in off-road transportation scenarios, such as mining, forestry, and sandy terrains, where vehicle mass can fluctuate significantly during loading and unloading operations, sometimes by as much as 30 tons \cite{pence2009sprung}. These variations lead to considerable changes in braking behavior, traction control, and rollover risk, making real-time estimation essential to maintain vehicle stability and safety under demanding conditions \cite{mcintyre2009two, pence2009sprung, yu2022mass}. In addition to safety considerations, accurate mass information improves energy efficiency by enabling better prediction of power demand and facilitating smoother velocity trajectories with reduced reliance on braking. It also supports high-level decision-making processes, including velocity planning, gear selection, and overload detection, all of which contribute to safer and more economical transport operations.  Moreover, accurate mass estimation is closely related to accurate powertrain modeling and control. Reflecting these practical demands, the development of reliable and real-time vehicle mass estimation techniques has attracted increasing attention in recent years~\cite{fathy2008online}.

A wide range of methods have been proposed for vehicle mass estimation, given its critical role in accurate longitudinal control and energy-efficient operation. These methods are generally categorized into ``gray box" and ``black box" approaches. ``Gray box" methods, which rely on vehicle longitudinal dynamics and are particularly suited for real-time implementation due to low computational complexity, include: i) Extended Kalman Filter (EKF)~\cite{zarringhalam2012comparative,lundin2012estimation,lingman2002road} and its extensions, such as integrations with model predictive control~\cite{winstead2005estimation}, adaptive filtering for real-time noise tuning~\cite{zhang2024identification,huang2014real}, anti-windup structures for handling outliers and poor excitation~\cite{altmannshofer2016robust}, and dual-filter frameworks for joint estimation of mass and inertia~\cite{hong2014novel,boada2019sensor}. ii) Least Squares (LS) based approaches, including Recursive Least-Squares (RLS) and Recursive Total Least-Squares (RTLS)~\cite{fathy2008online,rhode2012vehicle,lin2018method,chor2023robust,vahidi2005recursive,Koide:25}, are also widely applied. iii) Recursive estimation with polynomial chaos and signal-to-noise ratio based reliability~\cite{pence2013recursive}. “Black box” methods rely on large-scale sensor data and machine learning to infer vehicle mass from driving signals. Extensions include combining analytical models (e.g., EKF, RLS) with fuzzy logic or neural networks for enhanced adaptability~\cite{yu2022mass}, or integrating RLS with Long Short-Term Memory networks for improved accuracy in heavy-duty vehicles~\cite{isbitirici2025data}. Related efforts apply neural networks together with classical estimators for tasks such as slope and mass co-estimation in braking control~\cite{chen2023regenerative}, but these remain scenario-specific rather than addressing general real-time mass estimation. \cite{zhou2025hybrid} further proposed a hybrid model with physics-based estimation and neural traction modeling, but the approach relies on offline data collected under controlled tests, not online estimation during unconstrained transport.


Despite these developments, existing vehicle mass estimation methods typically rely on passively collected driving data and focus on improving estimator performance under given driving conditions. A similar pattern can be observed in broader automotive estimation problems, where studies on vehicle parameter identification, sideslip estimation, tire-force estimation, and damper-force estimation have significantly advanced parameter or state reconstruction, but usually under predefined driving conditions rather than through explicit design of the driving profile \cite{bevly2006integrating,antunes2019implementation,cordeiro2019estimation,pham2019design,pham2019real,pham2019unified}. In automotive system identification, informative excitation may be introduced through predefined maneuvers or standard input signals, such as swept-sine profiles, to improve parameter identifiability \cite{RIBEIRO2021104924}. However, such signals are usually selected to provide generally informative data, rather than optimized for a prescribed estimation accuracy. In contrast, optimal input design and application-oriented input design treat the excitation signal as a design variable and optimize it with respect to the intended estimation or control objective \cite{Ljung:99,honorio2018persistently,hjalmarsson2009system,annergren2017application}. Nevertheless, this input-design perspective has received limited attention in automotive mass estimation, where the driving profile itself is seldom optimized to achieve a prescribed estimation accuracy under practical vehicle constraints.

\begin{figure*}[!t]

    \begin{minipage}[l]{1.0\columnwidth}
        \centering
        \includegraphics[width=0.6\columnwidth]{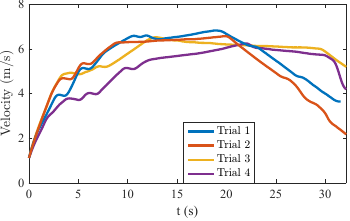}
  \caption{Normal driving velocity profiles of four trials.}
    \label{250625figure30}
    \end{minipage}
    \hfill{}
    \begin{minipage}[r]{1.0\columnwidth}
        \centering
     \includegraphics[width=0.6\columnwidth,]{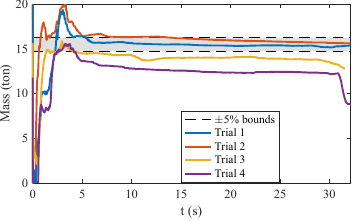}
  \caption{Least-squares mass estimation}.
    \label{250625figure33}
    \end{minipage}
    
\end{figure*}

Since mass estimation fundamentally relies on vehicle dynamic responses, the identifiability of mass parameters is closely related to the variability in acceleration, velocity, and traction force. Although the influence of input excitation on estimation performance has not been systematically studied, \cite{fathy2008online} suggests that informative data segments, typically those involving dynamic maneuvers such as acceleration or deceleration, tend to yield more accurate mass estimates. Nevertheless, most existing methods rely on passively identifying informative segments from recorded data, without explicitly addressing how input signals or driving profiles can be proactively designed to improve estimation accuracy. To illustrate the issue, we show in Fig.~\ref{250625figure30} the velocity profiles of four standard trajectories consisting of an acceleration segment followed by a retardation segment for the truck used in the experiments in this paper. These trajectories result in the mass estimates shown in Fig.~\ref{250625figure33} using a standard LS estimator, detailed later in this paper. 
For Trial 1 and Trial 2 in Fig.~\ref{250625figure30}, the LS estimate is close to the true mass, likely due to larger velocity variations and stronger acceleration phases in the corresponding profiles. However, such favorable excitation is not systematically guaranteed under normal driving conditions, as illustrated by Trials 3 and 4, which result in poor estimates in Fig.~\ref{250625figure33}. This highlights that achieving reliable estimation accuracy requires explicitly designed driving profiles rather than passively observed ones.

In this paper, we focus on the design of input signals, in the form of driving profiles, to improve estimation accuracy. The contribution lies in input design rather than in the development of a new estimation method. This perspective naturally leads to the question of how the excitation induced by a driving profile influences the identifiability of the mass parameter. This highlights that improving estimation accuracy requires not only better estimators, but also a systematic design of the excitation itself.

To address this question, we turn to input design, a concept rooted in system identification, where control inputs are deliberately optimized to enhance the information regarding target parameters \cite{Mehra:74,Goodwin&Payne:77,Gevers&Ljung:86,Ljung:99,Forssell&Ljung:00,Atkinson&Bailey:01,Gevers:11,Jansson:04a,Valenzuela:15a,pang2016data,bombois2006least}. In particular, the least-costly identification \cite{bombois2006least} and application-oriented input design (AOID) framework explicitly incorporates the intended estimation or control objective into the input signal design process \cite{hjalmarsson2009system,annergren2017application, larsson2014application,ebadat2017application, Hjalmarsson:04a, Annergren:12a}. Instead of focusing on the accuracy of the estimated parameters, this approach prioritizes excitations that are most relevant for the performance of the specific application in which the estimated model is to be used. This is typically achieved by analyzing how uncertainties in model parameters influence performance metrics, using tools such as experiment information matrices and application sensitivity analysis. In this paper, we mainly focus on time-domain feasibility, with explicit amplitude constraints on acceleration, velocity, and required accuracy to ensure physical realism. In the context of vehicle mass estimation, this framework offers a theoretically sound and practically viable strategy for generating driving profiles that are both informative and physically admissible, while meeting prescribed accuracy requirements. Although similar ideas have been explored in other domains \cite{Sigurdsson:24a, Rivera:09a, Lundh:24}, their targeted application to vehicle mass estimation under such physical constraints remains underdeveloped.

The main contributions of this paper are summarized as follows:

\begin{itemize}
\item[i)] Application-oriented problem formulation:  
This paper formulates vehicle mass estimation as an application-oriented system identification problem, where a confidence-bound constraint is imposed to guarantee that the estimation achieves the required accuracy.

\item[ii)] Constrained input design under realistic driving conditions:  
This paper develops input design strategies that account for physical constraints on acceleration, velocity, and excitation energy. The framework considers multiple objectives, including minimum-time, minimum-distance, and maximum estimation accuracy, and obtains driving profiles suitable for practical use and experimental planning.

\item[iii)] Global and rank-constrained methods for input design:
We adopt a global branch-and-bound method to solve the amplitude-constrained input design problem for moderately sized cases. 
Compared to prior approaches like~\cite{Sigurdsson:24a}, which rely on local optimization methods (e.g., Matlab’s {\tt fmincon}), the proposed methods offer more reliable global solutions.

\item[iv)] Analytical characterization of input structures:  
This paper provides theoretical insights into input design under acceleration and velocity constraints. For distance minimization, we identify a critical ratio between velocity and acceleration bounds that yields shorter distances than time-optimal designs. For time minimization, we construct periodic acceleration profiles and derive a sufficient condition on excitation length. We also show that the distance gap between the two objectives increases monotonically with the velocity range, highlighting the impact of input flexibility.

\item[v)] Phase-lag-free non-causal Wiener filtering via Empirical Bayes: A non-causal Wiener filter where the filter parameters are determined using data by way of the Empirical Bayes method is proposed. The method is applied in this paper for filtering accelerometer signals so as to avoid the phase-lag occurring with standard causal low-pass filters. 

\item[vi)] Validation through simulation and real-world experiments:
The proposed input design method is first evaluated in simulation under various constraints and objectives, demonstrating its ability to generate informative excitation. It is then validated on a Scania truck under two payload conditions. Results confirm that the designed trajectories meet the accuracy bounds and are practically feasible.

\item[vii)] Validation for RTLS: 
While the input design method uses the uncertainty of the least-squares estimate, using the Scania truck data, we show that the method also yields significant improvements when RTLS estimation is used. On these data it is also shown that Wiener filtering combined with least-squares is competitive with recursive total least-squares.

\end{itemize}



We stress that the main contribution of this paper is neither to propose a novel estimation method for a specific quantity such as mass, nor a novel longitudinal dynamics vehicle model. Instead, the objective is to develop a method for design of experiments for parameter estimation in longitudinal dynamics vehicle models. The result is a driving profile that ensures  a pre-specified estimation accuracy for a given target quantity. While the framework is demonstrated for vehicle mass estimation, it can adapted to other problems. To focus on the essentials a simple model is used, with details on how to handle more elaborate models being provided in Appendix \ref{modelapp}.

The remainder of the paper is organized as follows. Section~\ref{modelsec} introduces the vehicle longitudinal model, and Section~\ref{sec:mass} presents signal pretreatment and mass estimation formulation. Sections~\ref{sec:aoid} and \ref{sec:aoid_mass} develop and implement the AOID framework, respectively. Section~\ref{sec:opt} gives theoretical insights. Sections~\ref{sec:numerical} and \ref{sec:real_world} provide numerical and real-world experimental results, respectively. Section~\ref{sec:conclude} concludes the paper.

\section{Longitudinal Vehicle Dynamics Model}
\label{modelsec}
\begin{figure}[h]
    \centering
\includegraphics[width=0.45\linewidth]{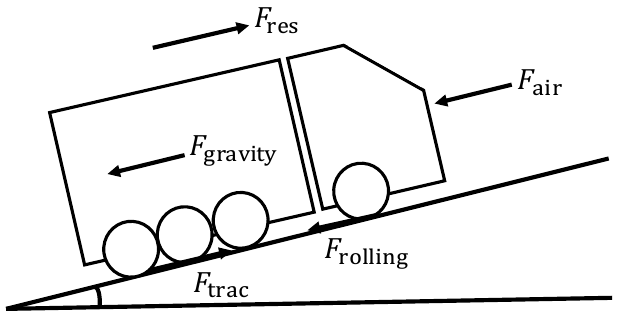}
    \caption{Longitudinal forces on the vehicle.}
    \label{fig:truck_draw}
\end{figure}

Consider a sampled data model, with $T_s$ denoting the sampling time. Sampling instances are denoted by $k$, e.g., $a(k)$ denotes the longitudinal acceleration at sampling instance $k$. 
A schematic of the forces acting on the vehicle in the longitudinal direction is given in Fig.~\ref{fig:truck_draw}. 

Here, $F_{\text{trac}}$ represents the effective traction force transmitted to the road surface, 
$F_{\text{air}}$ represents the air resistance, $F_{\text{rolling}}$ the rolling resistance and $F_{\text{gravity}}$ the gravitational force due to movements in a non-horizontal plane. The resultant force at sampling instance $k$ is
\begin{align}
\label{Fres}
      F_{\text{res}}(k):=F_{\text{trac}}(k)-F_{\text{air}}(k)-F_{\text{rolling}}(k)-F_{\text{gravity}}(k),
\end{align}
which is the force that changes the velocity of the truck. It should be noted that friction brakes are not included in the model since it is very difficult to estimate the force exerted by such brakes. Thus the model is valid only when friction brakes are not used. The relation between acceleration and the resulting force is given by Newton's second law. Since the acceleration is the result of the engine force, it may seem more natural from an input-output perspective to express this law as
\begin{align}
\label{newton2}
    a(k)=\frac{1}{m}\; F_{\text{res}}(k),
\end{align}
where $m$ is the mass of the truck, rather than the common expression
\begin{align}
\label{newton}
    F_{\text{res}}(k)=m \; a(k),
\end{align}
where the acceleration is given the interpretation as the input and the force as the output. However, there are several issues that need to be considered from a parameter estimation perspective. Firstly, while  \eqref{newton2} is a linear regression model with the inverse of the mass, $1/m$, as parameter, \eqref{newton} is a linear regression model with the quantity of interest $m$ as parameter. However, more important is to consider the impact of measurement errors. Commercial trucks include more or less sophisticated online estimators of the components of $ F_{\text{res}}(k)$. In practice, \(F_{\mathrm{res}}(k)\) is therefore not obtained from a single ideal sensor, but from a combination of sensors and online estimators, giving rise to modeling and estimation errors. With exact measurements of the acceleration $a(k)$ this would favor \eqref{newton} as model as additive output noise can be easily handled \cite{Ljung:99}.  However, the acceleration $a(k)$, measured by way of (inertial) accelerometers, can be subject to substantial measurement errors. This means that regardless of the choice of \eqref{newton2} or \eqref{newton} as model, we are faced with what is known as an errors-in-variables~(EIV) model where both inputs and outputs are subject to measurement errors. It is well-known that parameter estimation in such models are prone to systematic errors unless dedicated methods are used \cite{soderstrom2018errors}. 
One method that has been used for vehicle mass estimation is the total least-squares method \cite{Huffel:87}, see, e.g.,\cite{Rhode:12,Rhode:13} and more recently \cite{Koide:25}.

In practice, one may often avoid the use of EIV-methods, which can be quite involved, by pre-filtering the input signal. In a truck, the acceleration changes much slower than the noise in the accelerometers. Therefore low-pass filtering can reduce the problem of noisy accelerometer signals significantly. We will consider this approach in Section~\ref{filteringsec}.  

To focus on the problem of designing the driving profile for estimating the mass $m$, we consider $F_{\text{res}}(k)$ as being available. However, 
the driving profile design method developed below is applicable to more elaborate models. In Appendix \ref{modelapp} we provide more details on longitudinal vehicle dynamics modeling. We use the model in \cite{Koide:25}  to show how to adapt the proposed method to extended models, see also \cite{jessica2025}. In Section \ref{eivsec} we also provide real-world results when recursive RTLS is used as estimation method.


Thus our starting point is the model
 \begin{align}
\label{newton3}
    F_{\text{res}}(k)=m \; a(k)+e(k),
\end{align}
where $e(k)$ is the error in the estimate of the resulting force, which is assumed as a sequence of independent identically normal distributed random variables $e(k)\sim{\mathcal N}(0,\sigma_e^2)$. 

We also need to complement the model with the dynamics from the acceleration and retardation requests. The relation between these quantities comprises the whole drive-train and is thus rather complicated. Fortunately, a precise model for this is not necessary, as we see in the real-world validations of our method in Section~\ref{sec:real_world}. We assume that the acceleration responds in a first-order manner to change the accelerator and brake pedals, and model this  using a stable first-order transfer function with unity static gain
\begin{align}
\label{inputmodel}
    a(k)=F(q)u(k), \quad F(q)=\frac{(1-p)}{1-pq^{-1}},\quad p\in(0,1).
\end{align}
Here, the pole $p$ determines the time constant of the model. The model is adequate for the truck and operating conditions considered in the experiments in Section VIII. The truck was operated in a fixed gear and within a prescribed speed window, where the selected acceleration and deceleration levels were attainable after calibration. For another vehicle or operating region, this model may need to be re-identified using calibration data from the relevant speed, gear, and actuator range.

In summary, the considered framework consists of \eqref{newton3} and \eqref{inputmodel}, where $a(k)$ is obtained as a low-pass filtered version of the control input $u(k)$, which is physically realized by the accelerometer signal measuring the truck's longitudinal acceleration. The term $F_{\text{res}}(k)$ denotes an estimate of the resultant force on the truck, computed from the forces on the right-hand side of \eqref{Fres}. The pole $p$ in \eqref{inputmodel} is assumed to be identified {\em a priori} from  truck data, and is thus treated as known. Consequently, the only unknown parameter in the model is the mass $m$ in \eqref{newton3}.

\section{Pretreatment of signals and Mass Estimation}
\label{sec:mass}

From a parameter estimation perspective, the acceleration model \eqref{inputmodel} is not required, as the signal $\{a(k)\}$ is assumed to be directly available. However, the measured acceleration can be considerably noisy. An example from the truck used in the real-world experiments is shown by the blue line in Fig.~\ref{fig:zoom}(a). We next present a systematic approach for noise reduction, and address the problem of mass estimation based on model~\eqref{newton3}.


\subsection{Non-causal Wiener filtering of noisy signals}
\label{filteringsec}
As it is the noise from the accelerometer that is the primary concern in our application, we use this signal as an illustration. However, the results below can be applied to other signals as well. Consider therefore the simple model
\begin{align}
\label{astochmodel}
    a(k)=a_\circ(k)+e_a(k),
\end{align}
where $a_\circ(k)$ is the true acceleration and $\{e_a(k)\}$ is a sequence of independent identically distributed normal random variables $e_a(k)\sim{\mathcal N}(0,\sigma_a^2)$, modeling the measurement errors. By making the assumption that the true acceleration can be modeled as a stochastic process, we can use the non-causal Wiener filter to estimate $a_\circ(k)$ from $a(k)$. Denoting its power spectrum by $\Phi_{a_\circ a_\circ}(\omega)$, this filter has frequency function
\begin{align}
    W(e^{i\omega})=\frac{\Phi_{a_\circ a_\circ}(\omega)}{\Phi_{a_\circ a_\circ}(\omega)+\sigma_a^2},
    \label{wienerfilter}
\end{align}
see \cite{Kailath:2000}.  One can notice that phase-lag of this filter is zero, which is important as otherwise the acceleration and force will be out-of-sync due to the phase-lag introduced by the filter, leading to a bias in the LS estimate that will be presented in the next subsection. The performance of this filter depends critically on how well the used power spectrum $\Phi_{a_\circ a_\circ}(\omega)$ reflects the spectral properties of $a_\circ(k)$. Here, we consider a simple first-order dynamic model
\begin{align}
\label{aomodel}
    a_\circ(k)=F_a(q)v(k),\quad \text{where } F_a(q)=\frac{1}{1-\xi q^{-1}},
\end{align}
where $\{v(k)\}$ is a sequence of independent identically normal distributed random variables $v(k)\sim {\mathcal N}(0,\sigma_v^2)$. The pole $\xi$ determines the time constant of the model. This model gives
\begin{align}
    \Phi_{a_\circ a_\circ}(\omega)=|F_a(e^{i\omega})|^2\;\sigma_v^2.
    \label{firstordermodel}
\end{align}
\subsubsection{Estimating filter parameters}
The model \eqref{aomodel} with \eqref{firstordermodel} has two unknown parameters, the noise variance $\sigma_v^2$ and the filter pole $\xi$. One possibility to estimate these from data is to use the Empirical Bayes method \cite{Lehmann:98,RasmussenW:2006,PillonettoCD:2011} which simply amounts to maximum likelihood estimation. Assuming $N$ measurements to be available, let $a=[a(1), \cdots, a(N)]^T$, $v=[v(1),\cdots, v(N)]^T$ and $e_a=[e_a(1), \cdots,e_a(N)]^T$. Then using \eqref{aomodel}, equation \eqref{astochmodel} can be written on vector form as
\begin{align*}
    a=T(\xi)v+e_a\sim{\mathcal N}\left(0,T_a(\xi)T^T_a(\xi)\sigma_v^2+\sigma_{a}^2I\right)\nonumber,
\end{align*}
where $T_a(\xi)$ is a lower Toeplitz matrix with the first $N$ impulse response coefficients of $F_a(q)$ in the first column. We can now maximize the likelihood function for $a$ with respect to $\xi$, $\sigma_v^2$ and also $\sigma_a^2$. The negative log-likelihood function (omitting parameter independent terms) is given by
\begin{multline}
L(\xi,\sigma_v^2,\sigma_a^2)=a^T\left(T(\xi)T^T\xi)\sigma_v^2+\sigma_{a}^2I\right)^{-1}a\\
+\log \det \left(T(\xi)T^T\xi)\sigma_v^2+\sigma_{a}^2I \right)\nonumber,
\end{multline}
which, introducing $\kappa=\sigma_v^2/\sigma_a^2$, we can write as
\begin{multline}
L(\xi,\sigma_v^2,\sigma_a^2)=a^T\left(T(\xi)T^T\xi)\kappa+I\right)^{-1}a\;\frac{1}{\sigma_a^2}\\
+N\log\sigma_a^2+\log \det \left(T(\xi)T^T\xi)\kappa+I \right).
\label{marginal}
\end{multline}
It is straightforward to minimize this expression with respect to $\sigma_a^2$ by letting
\begin{align}
\label{sigmaa22}
    \sigma_a^2=\frac{1}{N}\; a^T\left(T(\xi)T^T\xi)\kappa+I\right)^{-1}a,
\end{align}
which inserted in \eqref{marginal} gives the condensed negative log-likelihood
\begin{multline}
    \tilde{L}(\xi,\kappa)=N-N\log N+N\log a^T\left(T(\xi)T^T\xi)\kappa+I\right)^{-1}a\\
    +\log \det \left(T(\xi)T^T\xi)\kappa+I \right)\nonumber.
\end{multline}
Minimizing this function numerically with respect to $\xi$ and $\kappa$ gives estimates $\hat{\xi}$ and $\hat{\kappa}$, from which also an estimate of $\sigma_a^2$ can be obtained from \eqref{sigmaa22} using $\xi=\hat{\xi}$ and $\kappa=\hat{\kappa}$.

\subsubsection{Realization of the Wiener filter} The filter \eqref{aomodel} with \eqref{firstordermodel} can be realized by a first-order filter 
\begin{align}
    F_W(q)=\frac{c}{1-\beta q^{-1}}\nonumber.
\end{align}
Using one forward and one backward filtering sweep as implemented in Matlab's {\tt filtfilt}. These operations give a filter with frequency function 
\begin{align}
  |F_W(e^{i\omega})|^2&=  \frac{c^2}{|1-\beta e^{-i\omega}|^2}=\frac{c^2}{1+\beta^2-2\beta \cos(\omega)}\nonumber \\
  &=\frac{c^2}{1+\beta^2}\; \frac{1}{1-\frac{2\beta}{1+\beta^2}\cos(\omega)}\nonumber.
\end{align}
The filter coefficients $c$ and $\beta$ can be determined by equating $|F_W(e^{i\omega})|^2$ with the frequency function for the Wiener filter~\eqref{wienerfilter}
\begin{align}
    W(e^{i\omega})&=\frac{\frac{\sigma_v^2}{|1-\xi e^{-i\omega}|^2}}{\frac{\sigma_v^2}{|1-\xi e^{-i\omega}|^2}+\sigma_a^2}
    \nonumber\\
    &=\frac{\sigma_v^2}{\sigma_v^2+\sigma_a^2(1+\xi^2)}\; \frac{1}{1-\frac{2\xi\sigma_a^2}{\sigma_v^2+\sigma_a^2(1+\xi^2)}\cos(\omega))}\nonumber.
\end{align}
This gives 
\begin{align}
\label{ceq}
\frac{c^2}{1+\beta^2}&=\frac{\sigma_v^2}{\sigma_v^2+\sigma_a^2(1+\xi^2)},\\
    \frac{2\beta}{1+\beta^2}&=\frac{2\xi\sigma_a^2}{\sigma_v^2+\sigma_a^2(1+\xi^2)}\nonumber.
\end{align}
Rewrite the second equation as a second-order equation in $\beta$
\begin{align}
    \beta^2-\frac{\sigma_v^2+\sigma_a^2(1+\xi^2)}{\xi\sigma_a^2}\beta+1=0\nonumber.
\end{align}
This equation shows that both two solutions are positive since their product and sum are both positive. Since the product of the solutions is 1, it has to hold that one solution is in the interval $(0,1)$ while the other is larger than 1. We are interested in the first solution since it corresponds to a stable filter. Thus $\beta$ is given by
\begin{align}
    \beta= \frac{\sigma_v^2+\sigma_a^2(1+\xi^2)}{2\xi\sigma_a^2}-\sqrt{\left(\frac{\sigma_v^2+\sigma_a^2(1+\xi^2)}{2\xi\sigma_a^2}\right)^2-1}\nonumber,
\end{align}
while $c$ is obtained from \eqref{ceq} as
\begin{align}
    c=\sqrt{(1+\beta^2)\frac{\sigma_v^2}{\sigma_v^2+\sigma_a^2(1+\xi^2)}}\nonumber.
\end{align}
\subsubsection{Real-data illustration} The Wiener filter is applied to estimate the true acceleration signal from noisy measurements. In this paper, the filter parameters are estimated as $\beta = 0.7092$ and $c = 0.2908$, and the corresponding signal-to-noise ratio is estimated to be 14~dB.

\begin{figure}[!t]
    \centering
    \begin{minipage}[b]{0.25\textwidth}
        \centering
        \includegraphics[width=\linewidth]{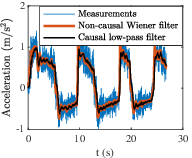}
        \caption*{(a) Measured/filtered signals}
    \end{minipage}\hfill
    \begin{minipage}[b]{0.234\textwidth}
        \centering
        \includegraphics[width=\linewidth]{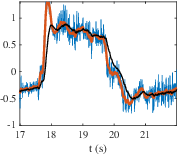}
        \caption*{(b) Zoom-in from (a)}
    \end{minipage}
    \caption{Measured acceleration and zoomed-in view.}
    \label{fig:zoom}
\end{figure}

Applying this method to the measured acceleration in Fig.~\ref{fig:zoom}(a) yields the red curve. 
For comparison, the signal obtained with a causal first-order low-pass filter, with unit static gain and a pole at 0.96, is shown in black. In contrast to the causal filter, which fails to capture the oscillations and introduces a visible time lag in the transients, particularly during the rapid rise of the third pulse, the Wiener filter accurately reproduces the oscillations. This difference is highlighted in Fig.~\ref{fig:zoom}(b), which provides a zoomed-in view of Fig.~\ref{fig:zoom}(a).


\subsection{Mass estimation}
\label{massestsec}

In this paper, the LS method is adopted for illustration purposes, due to its simplicity and well-understood properties. The proposed driving profile design framework is not tied to this specific estimation method and can be applied together with other estimation techniques after suitable modifications. We illustrate the impact driving profile design has on RTLS in Section \ref{eivsec}.
Neglecting the measurement errors in $a(k)$ (due to the low-pass filtering discussed in the previous subsection), \eqref{newton3} is amenable to LS estimation of $m$, giving the estimate 
\begin{align}
\hat{m}&=\frac{\sum_{k=1}^NF_{\text{res}}(k)a(k)}{R(N)}, \text{ where } R({N})=\sum_{k=1}^{{N}} a^2(k)
    \label{Rt}
\end{align}
Using \eqref{newton3}, we can express the estimate as
\begin{align}
    \hat{m}=m+\frac{\sum_{k=1}^N a(k)e(k)}{R(N)}\nonumber.
\end{align}

Thus the mass estimate is a Gaussian random variable and using the assumptions on the noise $e(k)$ we obtain 
\begin{align}
    \hat{m}\sim{\mathcal N} (m,P),\quad \text{where } P={\mathbb E}[(\hat{m}-m)^2]=\frac{\sigma_e^2}{R(N)}.
    \label{P}
\end{align}
From this, we see that $R(N)$, the sum of the squared accelerations, is the entity which influences the estimation accuracy and that can be influenced by the driving profile. 

\begin{figure}[!ht]
    \centering
\includegraphics[width=0.7\linewidth]{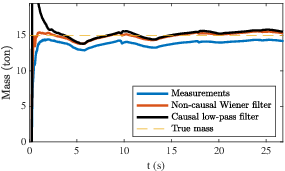}
    \caption{LS estimates of the mass using different regressors.}
    \label{lsfigure}
    
\end{figure}

\subsection{Real-world illustration}
Fig. \ref{lsfigure} plots the mass estimate versus sample size, with the acceleration signals in Fig. \ref{fig:zoom}(a) serving as regressors. The true mass is 15 tons, and clearly, using the raw accelerometer measurement results in a bias. The non-causal Wiener filter gives significantly better estimates for short sample sizes than the causal low-pass filter. The accuracy is also slightly improved. 


\section{Application-oriented input design}
\label{sec:aoid}
Below we outline the ideas behind AOID. Following this we consider input design for mass estimation in vehicles using this framework in the next section. Let $T$ denote the length of the experiment and  ${\mathcal U}_T$ denote the decision variables; typically, these are the available external excitation signal sequences to be applied during the experiment.

\subsection{Application-oriented model quality constraints}
As mentioned in Section I, it is important to know the vehicle mass in an autonomous heavy-duty vehicle during transport mission in order to achieve accurate powertrain control. In such a setting, we may let $J_{\text{app}}(m_\circ,m)$ be a function quantifying how the use of the mass $m$ in the control system of the truck degrades the fuel efficiency, as compared to when the control system has access to the true mass $m_\circ$. The primary objective of AOID is to design the input to the identification experiment such that the performance degradation caused by an estimated model $\hat{m}$ is acceptable from the point of view of the application, i.e., $J_{\text{app}}(m_\circ,\hat{m})$ is sufficiently small. 

More generally, let $\theta\in{\mathbb{R}}^n$ denote the unknown model parameters,  $J_{\text{app}}(\theta_\circ,\theta)$ denote the performance degradation when $\theta$ is used in the application instead of the true system parameters $\theta_\circ$. The set of acceptable models is then defined as
\begin{align}
    E_{\text{app}}:=\left\{\theta:\; J_{\text{app}}(\theta_\circ,\theta)\leq \frac{1}{2\gamma}\right\},
    \label{Eappp}
\end{align}
where $\gamma$ is a user chosen parameter known as the desired accuracy. The set of acceptable models thus represents all models that gives acceptable performance degradation when the model is used {\em in lieu} of the true system parameters. The input excitation is then designed to ensure that the estimated parameter vector $\hat{\theta} \in E_{\text{app}}$ with high probability. In general, this is a very difficult problem but via some approximations~\cite{hjalmarsson2009system,annergren2017application}, and this {\em application-oriented model quality constraint} can be (approximately) expressed as 
\begin{align}
    P^{-1}({\mathcal U}_T)\geq& \gamma\chi^2_\alpha(n)J_{\text{app}}''(\theta_\circ),
    \label{aomqc}
    \intertext{where}
    J_{\text{app}}''(\theta_\circ):=&\frac{\partial^2}{\partial \theta\partial\theta^T}J(\theta_\circ,\theta)\big|_{\theta=\theta_\circ}\nonumber.
\end{align}
$P({\mathcal U}_T)$ is the covariance matrix of the estimate $\hat{\theta}$ (which depends on the decision variables), and $\chi^2_\alpha(n)$ is the $\alpha$-percentile of the $\chi^2(n)$-distribution. The parameter $\alpha$ thus determines the desired probability. When the number of parameters is larger than one, this is thus a matrix inequality, and by a standard relaxation technique, the input power spectrum rather than the input sequence is considered as design variable \cite{Mehra:74,Goodwin&Payne:77,Jansson:04a}. Then, \eqref{aomqc} becomes a linear matrix inequality for linear models and thanks to this a range of AOID problems can be cast as semi-definite programs (SDPs), which can be solved reliably numerically, see \cite{hjalmarsson2009system,annergren2017application} for details and \cite{Annergren:12a} for an implementation. We can mention that there are also extensions of this technique able to handle non-linear models \cite{Valenzuela:15a}.

\subsection{Experimental constraints}
The excitation in an experiment needs to be constrained. Without such constraints, very large acceleration and velocity may result, as they yield a small covariance matrix $P$. The system to be modeled is subject to physical constraints, e.g., the power that can be generated by the engine is limited. We denote by ${\mathcal E}_T({\mathcal U}_T)$ the set of constraints for the experiment.

When the input power spectrum is chosen as the decision variable, the key to formulating the optimal input design problem as an SDP is that all constraints can be expressed in terms of the input power spectrum. Power constraints on the input excitation fit into this framework, whereas amplitude constraints on signals are not straightforward to use. In this paper we are faced with the latter type of constraints, as shown below.

\subsection{Experimental cost}
Let $V({\mathcal U}_T)$ denote a function quantifying the cost of an experiment as function of the decision variables. 

\subsection{Optimal input design formulations}
\label{oidsec}
This paper
focuses primarily on three performance metrics: experiment duration, experimental cost, and estimation accuracy. As introduced before, experiment duration is denoted by $T$ and estimation accuracy is determined by $\gamma$.
Let $V({\mathcal U}_T)$ denote a function quantifying the experimental cost of an experiment as function of the decision variables. Three types of optimal input design problems can be formulated as follows.
\begin{itemize}
    \item[i)] {\em Minimum experiment duration}
    \begin{align}
    \begin{split}
        \min_{{\mathcal U}_T}&\;T\\
        \text{subject to }&   P^{-1}({\mathcal U}_T)\geq \gamma\chi^2_\alpha(n) J_{\text{app}}''(\theta_\circ),\\
        &V({\mathcal U}_T)\leq c,\quad 
        {\mathcal E}_T({\mathcal U}_T).
        \end{split}
        \nonumber
    \end{align}
    In this formulation, the objective is to minimize the time it takes to achieve a certain accuracy $\gamma$, under the experimental constraints as well as a bound on the experimental cost. 
    \item[ii)] {\em Minimum experimental cost}
      \begin{align}
    \begin{split}
        \min_{{\mathcal U}_T}&\;V({\mathcal U}_T)\\
        \text{subject to }&   P^{-1}({\mathcal U}_T)\geq \gamma\chi^2_\alpha(n) J_{\text{app}}''(\theta_\circ),\;
        {\mathcal E}_T({\mathcal U}_T).
        \end{split}
        \nonumber
    \end{align}
     In this formulation, the objective is to minimize the experimental cost to achieve a certain accuracy $\gamma$, under the experimental constraints. 
     \item[iii)] {\em Maximum estimation accuracy}
      \begin{align}
    \begin{split}
        \max_{{\mathcal U}_T}&\;\gamma\\
        \text{subject to }&   P^{-1}({\mathcal U}_T)\geq \gamma\chi^2_\alpha(n) J_{\text{app}}''(\theta_\circ),\\
         &V({\mathcal U}_T)\leq c,\;
        {\mathcal E}_T({\mathcal U}_T).
        \end{split}
        \nonumber
    \end{align}
    In this formulation, the objective is to maximize the accuracy $\gamma$ under the experimental constraints as well as a bound on the experimental cost. 
\end{itemize}
 One may argue that the experimental time $T$ can be used as experimental cost, and thus i) could be merged into ii). However, as the number of decision variables depends on the experimental time, we treat these two cases separately. 

\section{AOID for mass estimation in vehicles}
\label{sec:aoid_mass}

The key objective of this section is to design the input (driving profile) such that the estimation accuracy requirement is satisfied. To achieve this we now apply AOID to our basic model \eqref{newton3}--\eqref{inputmodel}. The decision variables are the elements of the input sequence $\{u(k)\}$. With an experiment of $N$ samples, i.e., $T=T_sN$, we collect the corresponding input sequence in the column vector $u=[u(1), \cdots, u (N)]^T$. We thus have ${\mathcal U}_T=u$ for the mass estimation problem. The vehicle dynamics \eqref{inputmodel} can then be written in matrix form as 

\begin{align}
    a=Fu,
    \label{a}
\end{align}

\noindent where $F$ is a lower Toeplitz matrix with $[(1-p),  (1-p)p,  \cdots, (1-p)p^{N-1}]^T$ in the first column and $p$ denotes the pole of the acceleration dynamics model \eqref{inputmodel}.

\subsection{Application-oriented model quality constraint}
Among the various performance degradation measures $J_{\text{app}}$ relevant to mass estimation, we focus on accurate powertrain control and, for concreteness, consider the squared relative estimation error,

\begin{align}
\label{relerror}
    J_{\text{app}}(m_\circ,m)=\left(\frac{m-m_\circ}{m_\circ}\right)^2.
\end{align}
However, we emphasize it is just one possible choice, giving
\begin{align}
    J_{\text{app}}''(m_\circ)=2/m_\circ^2.
    \label{eq:J''}
\end{align}
The covariance $P$ is given by \eqref{P} and thus the quality constraint \eqref{aomqc} becomes
\begin{align}
    \sum_{k=1}^N a(k)^2\geq \frac{2\sigma_e^2 \gamma\chi_\alpha^2(1)}{m_\circ^2}.
        \label{qualityconstr}
\end{align}
Using \eqref{a}, we can express this as
\begin{align}
    u^TF^TFu\geq \frac{2\sigma_e^2 \gamma\chi_\alpha^2(1)}{m_\circ^2}\nonumber.
\end{align}

\subsection{Experimental constraints}
In a heavy-duty truck, the longitudinal acceleration and retardation are limited by the maximum engine torque and the retarder mechanism. Additional limitations arise from road slopes, such as uphill and downhill sections. These effects can be modeled by imposing distance-dependent constraints: let $d(k)$ be the distance traveled at sampling instant $k$, and let $a_{\min}(d)$ and $a_{\max}(d)$ denote the distance-varying bounds on acceleration. It is required that,
\begin{align}
    a_{\min}(d(k))\leq a(k)\leq a_{\max}(d(k))\nonumber.
\end{align}
Similarly, we can also impose constraints on the velocity $v(k)$
\begin{align}
    v_{\min}(d(k))\leq v(k)\leq v_{\max}(d(k))\nonumber.
\end{align}
In Appendix C, we show that the velocity and distance can be described as  
\begin{align}
    \hspace{-0.2cm}v(k) &= T_s \sum_{\ell=1}^k a(\ell), \label{velocity}\; 
    d(k) = T_s^2 \sum_{\ell=1}^{k} \frac{2(k-\ell)+1}{2}\; a(\ell). 
\end{align}
By stacking the values over the experiment horizon, we define $v=[v(1), \cdots,  v(N)]^T, d=[d(1), \cdots,  d(N)]^T.$
Using~\eqref{a}, the velocity and distance vectors can be written as
\begin{align}
    v &= T_s G F u, \quad
    d = T_s^2 H F u, \nonumber
\end{align}
where \( G \) is a lower Toeplitz matrix with ones in the first column, and \( H \in \mathbb{R}^{N \times N} \) is a lower triangular matrix of entries
\[
H_{k,\ell} =
\begin{cases}
k - \ell + \frac{1}{2}, & \text{if } \ell \leq k, \\
0, & \text{otherwise}.
\end{cases}
\]

 \subsection{Experimental costs}
A range of experimental costs $V$ can be associated with heavy-duty truck experiments, such as fuel consumption and wear-and-tear. In our mining site application, time is a key metric, but the traveled distance $d(N)$ may also be critical due to the limited road length $d_{\max}$. In this paper, the traveled distance can be interpreted as the experiment cost, as it is positively correlated with fuel consumption. In this paper, the travelled distance is interpreted as a proxy for the experiment cost, since it is positively correlated with fuel consumption.

\subsection{Basic AOID expressions}
Summarizing the above for the case that all of $a_{\min}$, $a_{\max}$,  $v_{\min}$ and $v_{\max}$  are constant, and the experimental cost is the distance traveled, gives the constraints
\begin{subequations}
\label{constraints}
\begin{align}
\label{constrainta}
   u^TF^TFu&\geq \frac{2\sigma_e^2 \gamma\chi_\alpha^2(1)}{m_\circ^2}, \\
   \label{constraintb}
   T_s^2H_{N\cdot}Fu&\leq d_{\max},\\
   \label{constraintc}
   a_{\min}{\mathbf 1}&\leq Fu\leq  a_{\max}{\mathbf 1},\\
   \label{constraintd}
    v_{\min}{\mathbf 1}&\leq T_s GFu\leq  v_{\max}{\mathbf 1},\\
      \label{constrainte}
    u_{\min}{\mathbf 1}&\leq u\leq  u_{\max}{\mathbf 1},
\end{align}
\end{subequations}
where ${\mathbf 1}$ denotes a vector of length $N$ with all elements being 1, and $H_{N\cdot}$ denotes the last row of $H$. Each of the optimal input design problems i)--iii) described above can be obtained from these expressions.

\subsection{Numerical considerations}
It is important to notice that while the last four constraints in~\eqref{constraints} are linear in the decision variable $u$, the first constraint is concave in $u$ and therefore all optimal input design problems are concave problems. Such problems are generally hard to solve numerically due to multiple local optima. In our setting, certain symmetry conditions allow one to easily deduce one or more optimal solutions, as discussed in Section \ref{sec:opt}. However, in general, one needs to use some numerical solvers. There is a wide range of alternatives, e.g., trust-region, sequential quadratic programming, and interior-point methods (all three available, e.g., in Matlab's {\tt fmincon}), relaxation techniques leading to SDPs, or global solvers, such as branch-and-bound (BnB) techniques. It is also possible to reformulate the contraints to enable SDP-relaxations as well as global optimization methods based on rank constraints, e.g. the Newton-based LMIRANK~\cite{Orsi:06}.
Below, we use the BnB-method {\tt bmibnb} available in YALMIP \cite{lofberg2004yalmip}, as this method has proven reliable for solving the type of optimal input design problems we consider for moderately sized problems. Due to the special structure of the constraints, it has also been shown to be meaningful to optimize specific input profiles~\cite{jessica2025}.

\subsection{The `chicken and egg' problem}

\label{sec:chicken}
It is common that optimal input design problems depend on the parameters that are going to be estimated. For the mass estimation problem, for example, \eqref{constrainta} depends on the unknown mass $m_\circ$. This can be traced back to that we are using the quadratic relative error in our performance degradation function $J_{\text{app}}$, see \eqref{relerror}. For the case of the quadratic error, $J_{\text{app}}(m_\circ,m)=(m-m_\circ)^2$, \eqref{constrainta} does not depend on $m_\circ$ and we thus have this problem for this choice of $J_{\text{app}}$. 

The most common approach to handle this is to use some \textit{a priori} estimates of the unknowns in the input design. Alternatively, one may use adaptive (sequential) input design where 
the parameters are estimated on-line and the optimal input design is updated at each time instance (or periodically) using the last available parameter estimate. It has been shown that such a procedure leads to the same asymptotic performance as if the true parameters had been used in the optimal input design in the first place \cite{Gerencser:07,Huang:14b}.

\section{Upper bounding the optimal solution}
\label{sec:opt}
In this section, we analyze some specific profiles to gain insights into the solution structure of optimal input design problems for mass estimation. For simplicity, the vehicle dynamics~\eqref{inputmodel} are neglected, hence the acceleration sequence $\{a(k)\}$ is considered as the decision variable instead of $\{u(k)\}$.


\subsection{Time minimization}

The quality constraint~\eqref{qualityconstr} can be expressed as
\begin{align}
R(N) \geq R_{\text{designed}}, \label{23}
\end{align}
where $R_{\text{designed}}= \frac{2\sigma_e^2 \gamma\chi_\alpha^2(1)}{m_\circ^2}.$ Equation \eqref{23} imposes a lower bound on $R(N)$. From the acceleration bounds it follows that
\begin{align}
R(N) \leq N \max \!\left( a_{\max}^2, a_{\min}^2 \right), \label{24}
\end{align}
providing an upper bound. Hence, $R(N)$ must lie within the interval defined by \eqref{23} and \eqref{24}. 
The upper bound in \eqref{24} is attained when the acceleration bounds are symmetric, i.e., 
$a_{\min} = -a_{\max}$, and the initial velocity is $v_{\min}$, provided that 
$T_s a_{\max} \leq v_{\max} - v_{\min}$. 
In this case, any input profile that alternates between $a_{\max}$ and $a_{\min}$ at every sample 
would respect the velocity limits and achieve the bound, resulting in a minimum-time design. A simple example is alternating between \( a_{\max} \) and \( a_{\min} \) at every sample. When vehicle dynamics~\eqref{inputmodel} cannot be neglected, such rapid switching becomes ineffective, as input changes may not immediately affect acceleration. A more realistic alternative is to apply maximum or minimum acceleration over longer intervals.

\begin{figure}[!ht]
    \centering
    \includegraphics[width=0.6\linewidth]{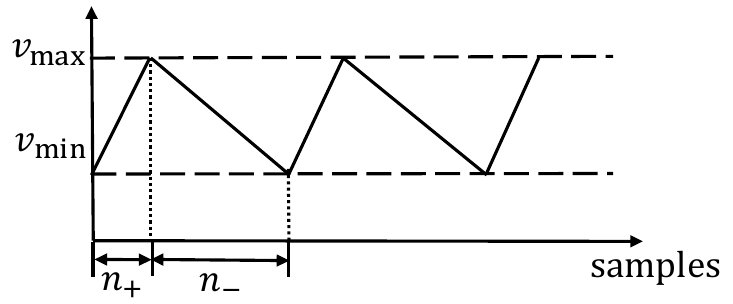}
    \caption{Schematic of the observed time-optimal structure.}
    \label{fig:time_s}
\end{figure}

Inspired by this observation, we consider the family of driving profiles shown in Fig.~\ref{fig:time_s}. Each acceleration phase uses the maximum acceleration \( a_{\max} \), and each deceleration phase uses the maximum deceleration \( a_{\min} \). In the symmetric case \( a_{\max} = -a_{\min} \), such profiles are optimal. The following result provides a sufficient condition on the number of cycles required to ensure feasibility, which also holds for the asymmetric case.

\begin{theorem}[Sufficient condition for feasibility via periodic acceleration]
\label{thm:periodic_feasibility}
Assume the system starts from initial velocity $ v_{\min}$, and $a_{\min} < 0 < a_{\max}$. Define: 
\[
n_+ := \left\lceil \frac{v_{\max} - v_{\min}}{T_s a_{\max}} \right\rceil, \qquad
n_- := \left\lceil \frac{v_{\max} - v_{\min}}{T_s |a_{\min}|}\right\rceil .
\]

Furthermore, define the admissible acceleration levels
\[
\bar a_+ := \frac{v_{\max}-v_{\min}}{T_s n_+},
\qquad
\bar a_- := -\frac{v_{\max}-v_{\min}}{T_s n_-}.
\]

Let $N = M(n_+ + n_-)$, where $M$ is the number of periods, and define the periodic acceleration pattern:
\[
a(k) =
\begin{cases}
\bar a_+, & \text{if } (k - 1) \bmod (n_+ + n_-) < n_+, \\
\bar a_-, & \text{otherwise}.
\end{cases}
\]
Then: 1) The velocity is within $[v_{\min}, v_{\max}]$; 2) The quality constraint $R(N) \geq R_{\text{designed}}$ is satisfied if the number of periods satisfies
    \[
    M \geq \frac{R_{\text{designed}}}{(n_+ \bar a_+^2 + n_- \bar a_-^2)}.
    \]

\end{theorem}
The proof is given in Appendix D. The theorem provides a constructive sufficient condition for feasibility of the time minimization problem under velocity and accuracy constraints. It yields an upper bound on the required horizon length \( N \) to satisfy the quality constraint. This bound does not guarantee optimality, and shorter feasible trajectories may exist outside this profile structure.


\begin{remark}
Returning to the symmetric case $a_{\max} = -a_{\min}$, let $n_c$ be a positive integer.
The profile reduces to an alternation between $a_{\max}$ and $-a_{\max}$, with each half cycle consisting of $n_c$ samples (thus $2n_c$ samples per full cycle).  
The velocity oscillates between $v_{\min}$ and $v_{\min} + n_c T_s a_{\max}$,  
and feasibility requires choosing
\begin{align}
    n_c \in \mathbb{Z}_{>0}, \qquad
    n_c \leq \left\lfloor \frac{v_{\max} - v_{\min}}{T_s a_{\max}} \right\rfloor .
    \nonumber
\end{align}
The excitation energy per cycle is \(2n_c a_{\max}^2\), and the quality constraint is satisfied if
\begin{align}
    M \geq \frac{R_{\mathrm{designed}}}{2n_c a_{\max}^2}.\nonumber
\end{align}
\end{remark}

\subsection{Distance minimization}
\label{sec:acc_rela}

\begin{figure}[!ht]
    \centering
 \includegraphics[width=0.6\linewidth]{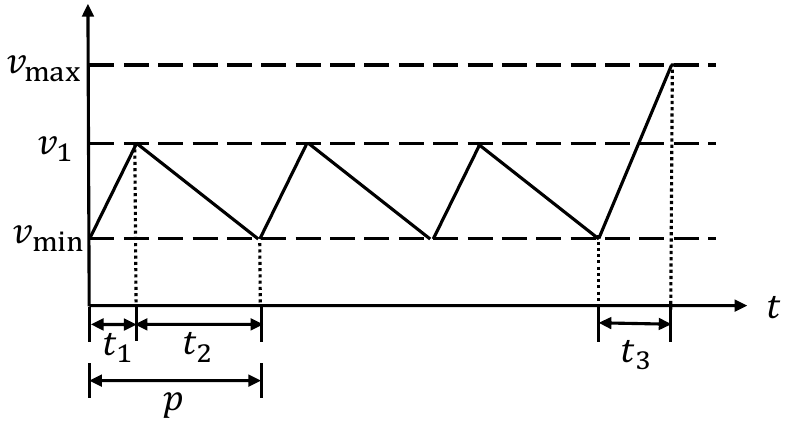}
    \caption{Schematic of the observed distance-optimal structure.}
    \label{fig:distance1}
\end{figure}

Numerical analysis reveals that, under fixed time and accuracy constraints, the optimal velocity profile for minimizing distance typically features repeated acceleration–deceleration cycles, ending with a final acceleration to \( v_{\max} \), as illustrated in Fig.~\ref{fig:distance1}. The vehicle accelerates at \( a_{\max} \) and decelerates at \( a_{\min} \) (assumed negative) during each cycle.

To gain analytical insight, we consider a simplified scenario where the vehicle starts from \( v_{\min} \), performs \( M \) repeated cycles between \( v_{\min} \) and an intermediate velocity \( v_1 \), and finally accelerates to \( v_{\max} \). Next, we turn to the case of distance minimization, for which a critical ratio between velocity and acceleration bounds can be established as follows.

\begin{theorem}[Critical Ratio for Distance Minimization]
\label{thm:distance-optimality}
For a fixed sampling period $T_s$, and time-invariant upper and lower bounds on acceleration and velocity ($a_{\max}$, $a_{\min}$, $v_{\max}$ and $v_{\min}$), use the profile in Fig. \ref{fig:distance1}, then the distance traveled to have $R(N)\geq R_{\text{designed}}$ is lower bounded by  
\begin{align}
\frac{T_sR_{\text{designed}} v_{\min}}{|a_{\min}| a_{\max}}  - \frac{(v_{\max} - v_{\min}) v_{\min}}{|a_{\min}|} + \frac{v_{\max}^2 - v_{\min}^2}{2 a_{\max}}.
\end{align}
The bound is achieved in the limit as $v_1\rightarrow v_{\min}$. For this limit case, and for given $a_{\min}$, $a_{\max}$ and $v_{\min}$, the distance is minimized by taking 
\begin{align}
\label{vmax}
    v_{\max}=\frac{a_{\max}}{|a_{\min}|}\, v_{\min},
\end{align}
which gives the distance
\begin{align}
\label{dstar}
d^\star =  \frac{T_sR_{\text{designed}} v_{\min}}{|a_{\min}| a_{\max}}
- \frac{(|a_{\min}|-a_{\max})^2v_{\min}^2}{2|a_{\min}|^2 a_{\max}}.
\end{align}
\end{theorem}
The proof is given in Appendix E. While this limiting case is not physically realizable, it can be approximated by taking \( v_1 \) close to \( v_{\min} \), and it offers valuable insights. First, it suggests that staying near the lower bound of the velocity range helps reduce the traveled distance.  
Second, expression~\eqref{dstar} shows that a smaller \( v_{\min} \) allows for a shorter distances.  
Third,~\eqref{vmax} reveals how \( v_{\max} \) should relate to \( a_{\max}/|a_{\min}| \) and \( v_{\min} \) to minimize distance.

\subsection{Monotonic growth of distance gap with velocity range}\label{subsec:vel_range}

The previous two theorems characterize the driving profiles for the minimum-time and minimum-distance objectives. Notably, the minimum-distance solution always yields a shorter travel distance than the minimum-time one, and this gap grows with increasing velocity range.

\begin{theorem}[Monotonic Amplification of the Travel Distance Gap with Respect to Velocity Range]
\label{thm:gap-monotone}
Denote the velocity gap by \(\Delta v=v_{\max}-v_{\min}\) and the distance gap by
\(
\Delta d(\Delta v)=d_{\mathrm{time}}-d_{\mathrm{distance}}
\), where 
\begin{align*}
    d_{\mathrm{time}}
    &=
    \frac{R_{\mathrm{designed}}T_s}{a_{\max}|a_{\min}|}
    \left(
        v_{\min}+\frac{\Delta v}{2}
    \right),
    \\
    d_{\mathrm{distance}}
    &=
    \frac{T_sR_{\mathrm{designed}}v_{\min}}{|a_{\min}|a_{\max}}
    -
    \frac{(v_{\max}-v_{\min})v_{\min}}{|a_{\min}|}
    \\
    &\quad
    +
    \frac{v_{\max}^2-v_{\min}^2}{2a_{\max}}.
\end{align*}

Then
\begin{enumerate}
\item[\textnormal{(i)}]  \(\displaystyle
      \Delta d(\Delta v)
    = -\,\frac{\Delta v^{2}}{2a_{\max}}
      + \frac{\Delta v}{|a_{\min}|}
        \Bigl(
           v_{\min} + \frac{T_sR_{\mathrm{designed}}}{2a_{\max}}-\frac{v_{\min} |a_{\min}|}{a_{\max}}
        \Bigr)\);
\item[\textnormal{(ii)}] Assuming that condition \eqref{vmax} holds, \( \Delta d(\Delta v) \) is \emph{strictly increasing} in \( \Delta v \).
\end{enumerate}
\end{theorem}
The proof is given in Appendix F. Hence, a larger velocity range produces
a larger travel distance gap between the minimum-time
and minimum-distance objectives.


\section{Numerical Examples}
\label{sec:numerical}

We set \( R_{\text{designed}} = 600 \) with the pole set as \( p = 0.979 \) in~\eqref{inputmodel}, matching the values used in the real-world experiments. This numerical study aims to verify whether the insights from Section~\ref{sec:opt} regarding the influence of velocity constraints on the minimum-time and minimum-distance profiles, derived from a restricted family of driving profiles, also hold for the optimal designs.

\begin{figure*}[!t]
    \centering
    \begin{minipage}[t]{0.24\textwidth}
        \centering
        \includegraphics[width=\linewidth]{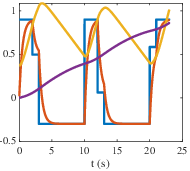}
        \caption*{(a) Small Vel. (Min. Time)}
    \end{minipage}
    \begin{minipage}[t]{0.24\textwidth}
        \centering
        \includegraphics[width=\linewidth]{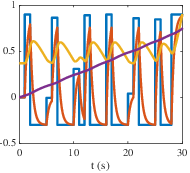}
        \caption*{(b) Small Vel. (Min. Dist.)}
    \end{minipage}
    \begin{minipage}[t]{0.24\textwidth}
        \centering
        \includegraphics[width=\linewidth]{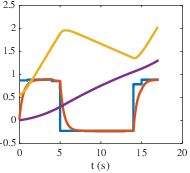}
        \caption*{(c) Large Vel. (Min. Time)}
    \end{minipage}
    \begin{minipage}[t]{0.24\textwidth}
        \centering
        \includegraphics[width=\linewidth]{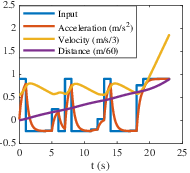}
        \caption*{(d) Large Vel. (Min. Dist.)}
    \end{minipage}
    \caption{Minimum-time and distance under small and large velocity ranges.}
    \label{fig:simulation}
\end{figure*}
\subsection{Effects of a small velocity range}

We set \(v_{\max} = {3.3\,\mathrm{m/s}}\), \(v_{\min} = {1.1\,\mathrm{m/s}}\), \(a_{\max} = 0.9\,\mathrm{m/s}^2\), and \(a_{\min} = -0.3\,\mathrm{m/s}^2\). Since the system dynamics in \eqref{inputmodel} have a static gain of 1, the control input \(u(k)\) is subject to the same bounds as the acceleration, i.e., \(u(k) \in [a_{\min}, a_{\max}]\). The optimization results for the minimum-time and minimum-distance problems under these settings are shown in Fig.~\ref{fig:simulation} (a) and (b), respectively. To enhance visual clarity, the velocity and distance are scaled by factors of $1/10$ and $1/60$, respectively, as indicated in the legend. The blue curve represents the input profile \(u(k)\), corresponding to the accelerator and brake actions. The red curve shows the acceleration profile \(a(k)\), which lags behind \(u(k)\) due to the system dynamics in \eqref{inputmodel}. The yellow and purple curves depict the velocity \(v(k)\) and distance \(d(k)\), respectively.

The optimal solution to the minimum-time problem results in a travel time of 23 seconds and a total distance of 52.3\,m. By extending the time horizon by 7 seconds in the minimum-distance optimization, the resulting trajectory achieves a shorter travel distance of 46.3\,m, corresponding to a reduction of approximately 6\,m, or about 13\% compared to the minimum-time case.


\subsection{Effects of a large velocity range}

We set \(v_{\max} = {6.4\,\mathrm{m/s}}\), \(v_{\min} = {1.1\,\mathrm{m/s}}\), and \(a_{\max} = 0.9\,\mathrm{m/s}^2\), and \(a_{\min} = -0.23\,\mathrm{m/s}^2\). The optimization results for the minimum-time and minimum-distance under these settings are shown in Fig.~\ref{fig:simulation} (c) and (d), respectively. The optimal solution to the minimum-time problem results in a travel time of 17 seconds and a distance of 77.8\,m. When the time horizon is extended by 6 seconds in the minimum-distance design, the resulting trajectory reduces the travel distance to 53.6\,m, a decrease of approximately 24\,m or 30\%. This reduction is notably greater than what is typically observed under a smaller velocity range.


\subsection{Summary}

The simulation results consistently align with the theoretical acceleration-deceleration patterns. For the minimum-time objective, the system accelerates rapidly to \(v_{\max}\) before decelerating to \(v_{\min}\), while for the minimum-distance objective, the velocity remains low initially and gradually increases toward \(v_{\max}\) near the end. This behavior is consistent with the structural patterns shown in Fig.~\ref{fig:time_s} and Fig.~\ref{fig:distance1} and discussed in Section~\ref{sec:opt}.

Moreover, the total travel distance under the minimum-distance objective is consistently shorter than that under the minimum-time objective, and this difference becomes more pronounced as the velocity range \(\Delta v = v_{\max} - v_{\min}\) increases, which is consistent with Theorem~3.

\section{Real World Results}
\label{sec:real_world}
\begin{figure}[!ht]
    \centering
    \begin{minipage}{0.49\columnwidth}
        \centering
        \includegraphics[width=\linewidth,height=3cm]{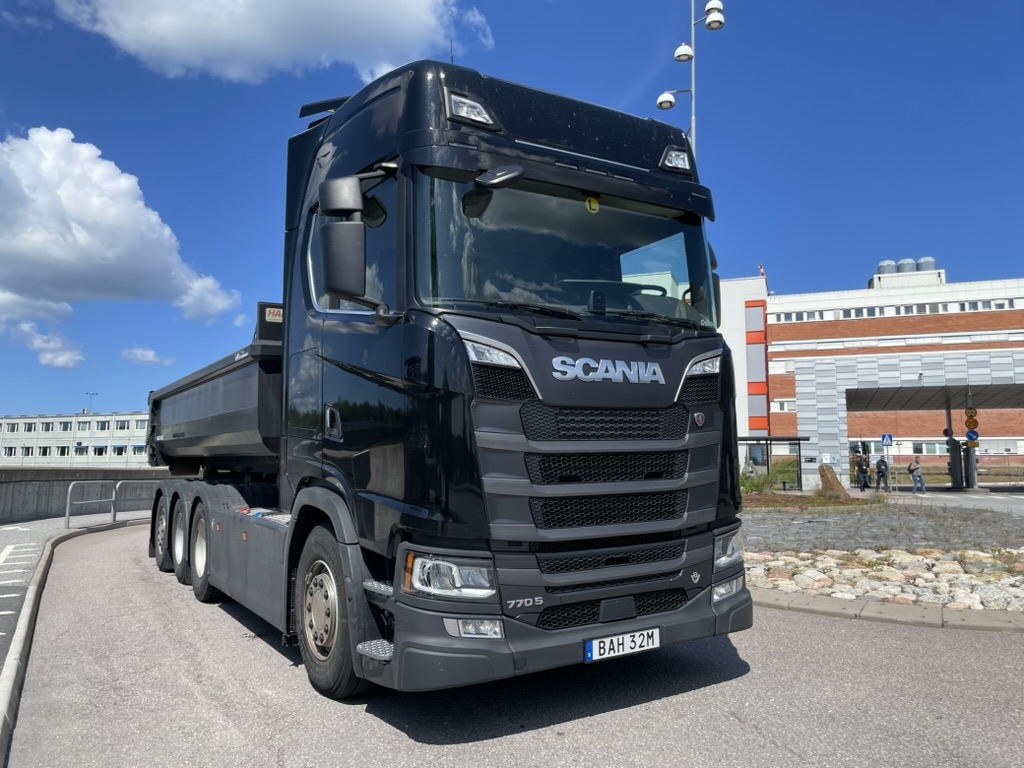}
        \caption*{(a) Test truck}
        \label{fig:truck}
    \end{minipage}
    \hfill
    \begin{minipage}{0.49\columnwidth}
        \centering
        \includegraphics[width=\linewidth,height=3cm]{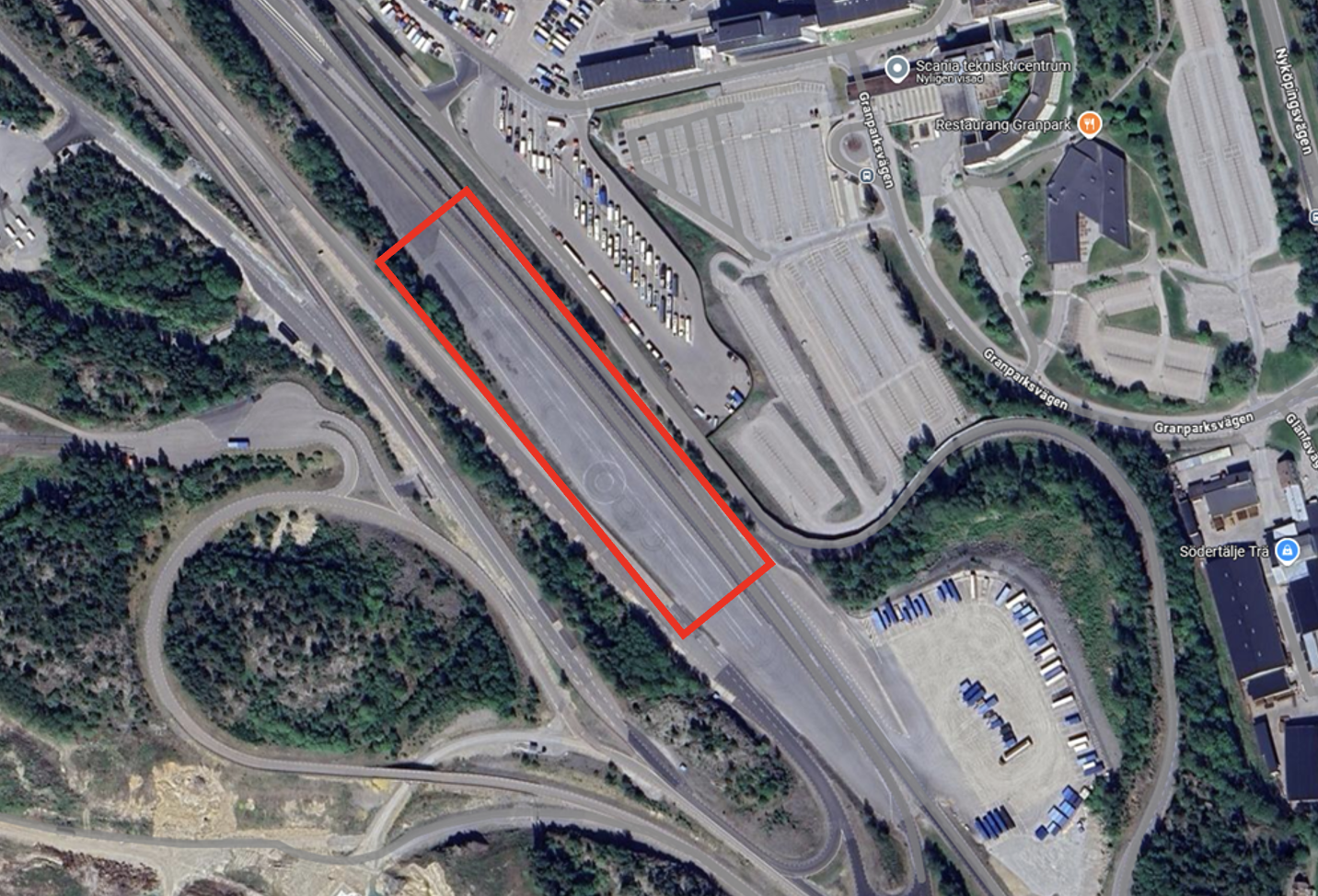}
        \caption*{(b) Test road}
        \label{fig:road}
    \end{minipage}
    \caption{Test truck and test road at Scania’s site in Stockholm.}
    \label{fig:truck_and_road}
\end{figure}

We implemented the proposed method on a Scania heavy-duty rigid chassis S-Cab truck to collect data under realistic conditions. The truck has a 770hp diesel engine, air suspension, 4 axles (1 front axle, 2 drive axles with twin montage, and 1 support axle), and is equipped with automated manual transmission and exhaust brake and retarder. The total mass could be manipulated through the use of different containers filled with gravel. The experiments were carried out on a flat test track with negligible road slope using production-grade sensors and data acquisition systems. The tests were conducted under stable weather conditions with minimal wind and were performed in both driving directions to mitigate potential bias caused by residual slope or wind effects. These controlled experimental conditions ensure that the collected data are reliable, that environmental disturbances are minimized, and that the designed excitation profiles can be safely executed within normal operating limits of the vehicle without introducing unintended disturbances. Fig.~\ref{fig:truck_and_road} shows the test vehicle and the road section at Scania’s test site in Stockholm, where all experiments took place.

\subsection{Set-up}
\label{subsec:setup}
The measured mass of the truck is $15.01$ tons for the first payload and $31.28$ tons for the second payload. However, in addition to this, the inertia of the 12 rotating wheels has to be accounted for. This contribution is estimated to be 528~kg. Thus, the true parameter for the two cases are $m_\circ^1=15.01+0.528=15.5$ tons and $m_\circ^2=31.28+0.528=31.8$ tons, respectively. Based on previous trials, the pole $p$ in the input dynamics \eqref{inputmodel} is set to $p=0.979$. In order to obtain a large difference in the traveled distance for the minimum-time and the minimum-distance design, the velocity range is set to $v_{\min}=1.7$ m/s and $v_{\max}=6.4$ m/s. 
The maximum acceleration is set to $a_{\max} = 0.9\,\mathrm{m/s}^2$, and a minimum acceleration of $a_{\min} = -0.4\,\mathrm{m/s}^2$ is chosen. No gear shift is allowed during the experiment time, and the experiment is considered to begin when the truck reaches the minimum velocity for the first time. A human driver is present, but the driving profile is executed by custom in-truck software, which controls the accelerator-pedal and auxiliary-brake-lever signals during acceleration and deceleration segments. The designed driving profile was implemented by a simple event-based procedure rather than by a dedicated feedback tracking controller. For each acceleration or deceleration segment, the software applied a constant accelerator-pedal percentage or auxiliary-brake-lever percentage, selected from calibration runs, until a specified time had elapsed, a target velocity was reached, or a prescribed velocity limit was reached. Therefore, the implemented profile was not expected to track the designed profile exactly. Deviations and overshoot can occur, especially when switching between propulsive and braking force, due to drivetrain and actuator dynamics.

We use the model \eqref{newton3} complemented with an offset term
\begin{align}
    F_{\text{res}}(k)=m\,a(k)+\delta+e(k),
    \label{eq:fres}
\end{align}
where the offset $\delta$ is estimated together with the mass $m$. Thus $n=2$ parameters are estimated and the procedure in Section~\ref{sec:nuis} is used to handle this. 

For the model quality constraint \eqref{constrainta}, the 99\%-percentile ($\alpha=0.99$) is used, giving $\chi_\alpha^2(2)=9.2$. The constraint is expressed as 

\begin{align}
    u^TF^TFu=\sum_{k=1}^N a(k)^2 \geq R_{\text{designed}}\nonumber,
\end{align}

\noindent {with, see \eqref{constrainta}, $R_{\text{designed}}=\frac{2\sigma_e^2\gamma \chi_\alpha^2(2)}{m_\circ^2}$. 
From \eqref{Eappp} we have that the upper bound on the squared relative error $|\hat{m}-m_\circ|^2/m_\circ^2$ for which the design is made is $1/(2\gamma)$, and hence a desired relative error of $3.9\%$ for the mass estimate gives $\gamma=1/(2\, \cdot\,  0.039^2)$. Assuming that the noise variance is $\sigma_e^2 = 0.1\, m_\circ^2$, gives $R_{\text{designed}}\approx 600$ which is the lower bound used in \eqref{constrainta}.}

In the plots for the mass estimates below (Figs. 11 and 12), the bounds $(1\pm \text{Relative Error}_{\text{designed}})m_\circ$ are shown as dashed lines. The noise variance used in this expression is taken as the mean of the noise variance estimates from the different experiments using the same input design. 

\begin{figure}[t]
    \centering
    \includegraphics[width=0.5\columnwidth]{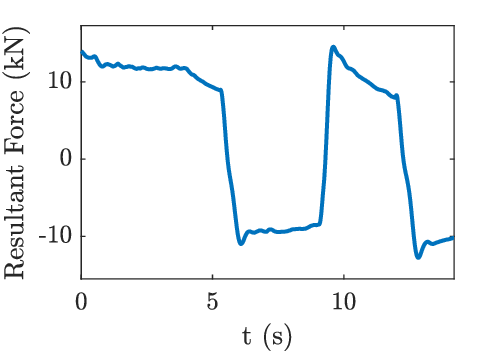}
    \caption{Resultant force signal used in the LS mass estimation from one minimum-time test run with the 15 tons payload.}
    \label{fig:Fres_15t_mintime}
\end{figure}

\begin{figure*}[!ht]
    \centering

    \begin{minipage}{0.24\textwidth}
        \centering
        \includegraphics[width=\linewidth]{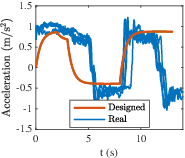}
        \caption*{(a) Acceleration (Min. Time)}
    \end{minipage}
    \hfill
    \begin{minipage}{0.24\textwidth}
        \centering
        \includegraphics[width=\linewidth]{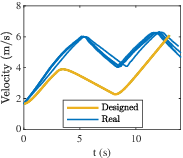}
        \caption*{(b) Velocity (Min. Time)}
    \end{minipage}
    \hfill
    \begin{minipage}{0.24\textwidth}
        \centering
        \includegraphics[width=\linewidth]{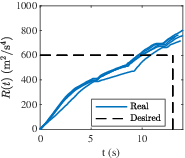}
        \caption*{(c) $R(t)$ (Min. Time)}
    \end{minipage}
    \hfill
    \begin{minipage}{0.24\textwidth}
        \centering
        \includegraphics[width=\linewidth]{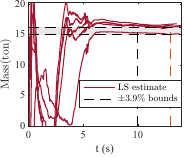}
        \caption*{(d) Mass Estimate (Min. Time)}
    \end{minipage}


    \begin{minipage}{0.24\textwidth}
        \centering
        \includegraphics[width=\linewidth]{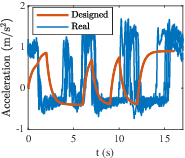}
        \caption*{(e) Acceleration (Min. Dist.)}
    \end{minipage}
    \hfill
    \begin{minipage}{0.24\textwidth}
        \centering        \includegraphics[width=\linewidth]{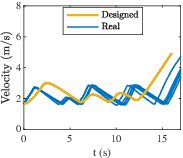}
        \caption*{(f) Velocity (Min. Dist.)}
    \end{minipage}
    \hfill
    \begin{minipage}{0.24\textwidth}
        \centering
        \includegraphics[width=\linewidth]{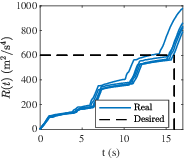}
        \caption*{(g) $R(t)$ (Min. Dist.)}
    \end{minipage}
    \hfill
    \begin{minipage}{0.24\textwidth}
        \centering
        \includegraphics[width=\linewidth]{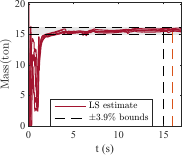}
        \caption*{(h) Mass Estimate (Min. Dist.)}
    \end{minipage}
    \caption{Acceleration, velocity, $R(t)$, and mass estimate for the minimum-time and distance designs ($m_\circ = 15.5$ tons).}
    \label{fig:min_time_distance_all}
    
\end{figure*}

\begin{figure*}[ht]
    \centering

    \begin{minipage}{0.24\textwidth}
        \centering
        \includegraphics[width=\linewidth]{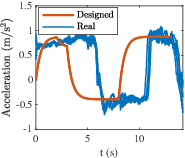}
        \caption*{(a) Acceleration (Min. Time)}
    \end{minipage}
    \hfill
    \begin{minipage}{0.24\textwidth}
        \centering        \includegraphics[width=\linewidth]{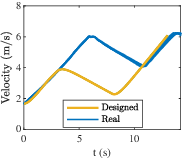}
        \caption*{(b) Velocity (Min. Time)}
    \end{minipage}
    \hfill
    \begin{minipage}{0.24\textwidth}
        \centering
        \includegraphics[width=\linewidth]{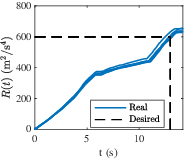}
        \caption*{(c) $R(t)$ (Min. Time)}
    \end{minipage}
    \hfill
    \begin{minipage}{0.24\textwidth}
        \centering
        \includegraphics[width=\linewidth]{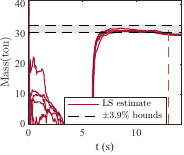}
        \caption*{(d) Mass Estimate (Min. Time)}
    \end{minipage}


    \begin{minipage}{0.24\textwidth}
        \centering
        \includegraphics[width=\linewidth]{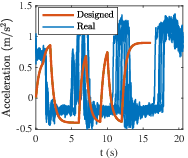}
        \caption*{(e) Acceleration (Min. Dist.)}
    \end{minipage}
    \hfill
    \begin{minipage}{0.24\textwidth}
        \centering
        \includegraphics[width=\linewidth]{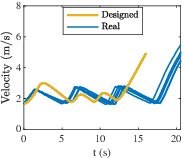}
        \caption*{(f) Velocity (Min. Dist.)}
    \end{minipage}
    \hfill
    \begin{minipage}{0.24\textwidth}
        \centering
        \includegraphics[width=\linewidth]{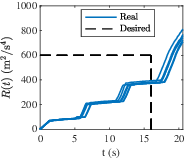}
        \caption*{(g) $R(t)$ (Min. Dist.)}
    \end{minipage}
    \hfill
    \begin{minipage}{0.24\textwidth}
        \centering
        \includegraphics[width=\linewidth]{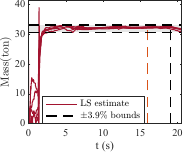}
        \caption*{(h) Mass Estimate (Min. Dist.)}
    \end{minipage}

    \caption{Acceleration, velocity, $R(t)$, and mass estimate for the minimum-time and distance designs ($m_\circ = 31.8$ tons).}
    \label{fig:design_results_all}
\end{figure*}

\subsection{Results for the 15 tons payload}
\subsubsection{Minimum-time design} The minimum-time design yielded a total duration of $T = 13$~s. Before presenting the experimental results, Fig.~\ref{fig:Fres_15t_mintime} shows the resultant force signal $F_{\text{res}}(k)$ from one representative test run with the 15-ton payload. This signal is used together with the measured longitudinal acceleration in the LS regression model \eqref{eq:fres}. Five experiments are conducted, with results shown in the upper panel of Fig.~\ref{fig:min_time_distance_all}. As seen in Fig.~\ref{fig:min_time_distance_all}(a), the simple control system (see Section~\ref{subsec:setup}) struggled to track the designed acceleration profile, leading to deviations in the resulting velocity profile (Fig.~\ref{fig:min_time_distance_all}(b)). Consequently, the designed value $R_{\text{designed}}$ is reached earlier than planned, around 10--11~s, as shown in Fig.~\ref{fig:min_time_distance_all}(c), where the dashed line marks the target duration $T = 13$~s. Here, \(t=N T_s\). Fig.~\ref{fig:min_time_distance_all}(d) illustrates how the LS mass estimates evolve across the experiments. The black vertical line indicates the average actual time when $R_{\text{designed}}$ is reached, while the red line shows the designed time. Although most estimates converge within a band comparable to $2\text{Relative Error}_{\text{designed}}$, a slight bias appears to be present.




\subsubsection{Minimum-distance design}
For the minimum-distance design, a total duration of $T = 16$~s is used, allowing 3 additional seconds to reach $R_{\text{designed}}$. Five experiments are conducted, with results presented in the lower panel of Fig.~\ref{fig:min_time_distance_all}. As shown in Fig.~\ref{fig:min_time_distance_all}(e), the control system again struggled to follow the designed acceleration profile. Nevertheless, the resulting velocity profile retained a shape similar to the desired one, as illustrated in Fig.~\ref{fig:min_time_distance_all}(f). As a result, $R_{\text{designed}}$ is reached in close alignment with the design specifications, as seen in Fig.~\ref{fig:min_time_distance_all}(g), where the dashed vertical line marks the end of the experiment ($T = 16$~s). Fig.~\ref{fig:min_time_distance_all}(h) shows the evolution of the LS mass estimates over time across the five experiments. The black vertical line indicates the average time when $R_{\text{designed}}$ is reached, while the red line indicates the designed time. It can be observed that the estimates converge to the desired relative accuracy within the allotted experiment duration.



\subsubsection{Comparison of the designs}
\label{comp15tonsec}
\begin{figure*}[!t]
    \centering

    \begin{minipage}{0.24\textwidth}
        \centering
        \includegraphics[width=\linewidth]{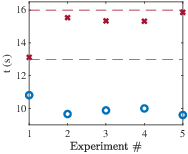}
        \caption*{(a) Min. Time  (15.5 tons)}
    \end{minipage}
    \hfill
    \begin{minipage}{0.24\textwidth}
        \centering
        \includegraphics[width=\linewidth]{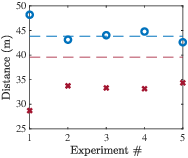}
        \caption*{(b) Min. Dist. (15.5 tons)}
    \end{minipage}
    \hfill
    \begin{minipage}{0.24\textwidth}
        \centering
        \includegraphics[width=\linewidth]{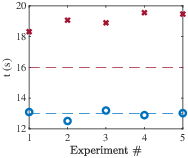}
        \caption*{(c) Min. Time (31.8 tons)}
    \end{minipage}
    \hfill
    \begin{minipage}{0.24\textwidth}
        \centering
        \includegraphics[width=\linewidth]{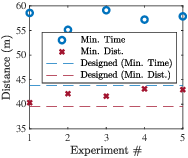}
        \caption*{(d) Min. Dist. (31.8 tons)}
    \end{minipage}

    \caption{Time and distance to reach $R_{\text{designed}}$ under different mass configurations. Dashed lines indicate the designed values.}
    \label{fig:reach_r_all}
    
\end{figure*}

Fig.~\ref{fig:reach_r_all}(a) shows the time it takes to reach $R_{\text{designed}}$ for the different trials and the different designs. We can see that the actual times match the designed ones relatively well, considering the challenges to follow the designs. As shown in Fig.~\ref{fig:reach_r_all}(b), the same goes for the distances it takes to reach $R_{\text{designed}}$.

In particular, the minimum-time design systematically leads to experiments requiring shorter time than the minimum-distance designs, despite the difficulties in following the trajectory. It takes an average of 10 s and 15 s to reach the design objective for the minimum-time and minimum-distance designs, respectively. 

Conversely, the minimum-distance design systematically leads to experiments requiring shorter distances to be traveled than the minimum-time designs. It takes an average of 32.7~m and 44.6 m to reach the design objective for the minimum-time and minimum-distance designs, respectively.



\subsection{Results for 31 tons payload}
\subsubsection{Minimum-time design}
The minimum-time design resulted in a total duration of $T=13$~s. Five experiments are conducted, with results shown in the upper panel of Fig.~\ref{fig:design_results_all}. As seen in Fig.~\ref{fig:design_results_all}(a), the achieved acceleration profile roughly follows the desired shape. However, prolonged high acceleration led to higher-than-expected velocities midway through the trials, see Fig.~\ref{fig:design_results_all}(b). Still, due to the overall similarity in acceleration shape, $R_{\text{designed}}$ is reached at the intended time, as shown in Fig.~\ref{fig:design_results_all}(c), where the dashed line marks the designed end time $T=13$~s. Fig.~\ref{fig:design_results_all}(d) presents the LS mass estimates over time for the five trials. The black vertical line indicates the average time $R_{\text{designed}}$ is reached, while the red line shows the designed target time. The estimates converge to a strip of width approximately equal to $\text{Relative Error}_{\text{designed}}$, except for one outlier, though a slight bias appears to be present.



\subsubsection{Minimum-distance design}
For the minimum-distance design, $T = 16$~s is used, i.e., 3 extra seconds are allowed to reach $R_{\text{designed}}$. Five experiments are conducted, as shown in the lower panel of Fig.~\ref{fig:design_results_all}. As seen in Fig.~\ref{fig:design_results_all}(e), the acceleration profile is mostly followed, except that the final large acceleration occurred too late. This is also reflected in the velocity profiles in Fig.~\ref{fig:design_results_all}(f). Consequently, $R_{\text{designed}}$ is reached several seconds after the target time, as shown in Fig.~\ref{fig:design_results_all}(g). Despite this, the LS mass estimates in Fig.~\ref{fig:design_results_all}(h) reached the desired accuracy well before $R_{\text{designed}}$ is actually achieved (around 19 s, marked by the black vertical line). 



\subsubsection{Comparison of the designs}
From Fig.~\ref{fig:reach_r_all}(c) and (d), we can draw the same conclusions as we did for the 15 tons payload case, see Section \ref{comp15tonsec}.
That is the actual times and distances required to reach $R_{\text{designed}}$
match the designed ones relatively well. Furthermore, the minimum-time design systematically leads to shorter experiments whereas the minimum-distance design systematically leads to shorter distances traveled. It takes an average of 12 s and 19 s to reach the design objective for the minimum-time and minimum-distance designs, respectively. It takes an average of 57.6 m and 42.0 m to reach the design objective for the minimum-time and minimum-distance designs, respectively. 


\subsection{Quantitative performance evaluation}
To quantitatively evaluate the estimation performance, both the Root Mean Square Error (RMSE) and the Normalized Mean Square Error (NMSE) are adopted. 
For each experimental run $i$, the estimation error is evaluated at the time instant $t_i^*$ when the desired accuracy level is achieved. 
The estimation error at this time instant is defined as

\[
e_i = \hat{m}_i(t_i^*) - m_{\circ},
\]

\noindent where $\hat{m}_i(t_i^*)$ denotes the estimated mass at time $t_i^*$. 
The RMSE and NMSE across all experimental runs ($N_{\mathrm{exp}}=5$ for each mass and design objective) are computed as
\[
\mathrm{RMSE} = \sqrt{\frac{1}{N_{\mathrm{exp}}} \sum_{i=1}^{N_{\mathrm{exp}}} e_i^2},\quad  \mathrm{NMSE} =  \mathrm{RMSE}^2/m_{\circ}^2.
\]
The NMSE provides a scale-invariant performance metric, which is particularly useful when comparing estimation accuracy across different true mass values. The NMSE can also be interpreted as the squared RMSE normalized by the true mass.
The quantitative performance under different masses and design objectives is summarized in TABLE~\ref{tab:performance}. The results show that the estimation errors remain small across all cases, and that the minimum-distance design generally achieves lower errors than the minimum-time design.
\begin{table}[h]
\centering
\caption{Quantitative performance in terms of RMSE and NMSE under different masses and design objectives.}
\label{tab:performance}
\renewcommand{\arraystretch}{1.3}
\setlength{\tabcolsep}{6pt}
\begin{tabular}{c c c c}
\hline
Mass ($10^3$ kg) & Objective & RMSE & NMSE \\
\hline
15.5 & Min. time  & $9.36 \times 10^{-1}$ & $3.65 \times 10^{-3}$ \\
15.5 & Min. dist. & $1.30 \times 10^{-1}$ & $7.02 \times 10^{-5}$ \\
31.8 & Min. time  & $7.33 \times 10^{-1}$ & $5.32 \times 10^{-4}$ \\
31.8 & Min. dist. & $3.15 \times 10^{-1}$ & $9.84 \times 10^{-5}$ \\
\hline
\end{tabular}
\end{table}


\subsection{Results using recursive total least-squares}
\label{eivsec}
We will now examine the effect driving profile design has on the estimation accuracy when the recursive total least-squares method presented in \cite{Koide:26} is used for estimating the mass. Being a development of the recursive total least-squares method presented in \cite{Koide:25}, this method is based on sequential quadratic programming and has been used in the thesis of Koide \cite{Koidethesis:26} for mass estimation in vehicles. 

Our implementation follows Algorithm 3 in \cite{Koidethesis:26}, with the additional feature that noise covariances are estimated as in Algorithm 4 in the same reference\footnote{It should be noted that it is well known that the noise covariances cannot be consistently estimated, the ratio between the output noise and the noise on the regressors needs to be known \cite{soderstrom2018errors}. However, the results in \cite{Koidethesis:26} indicate that it is indeed beneficial for the estimation accuracy to estimate these covariances.}. For initial values and parameter settings we use the values in Table 4.4 in \cite{Koidethesis:26}, save for the forgetting factor $\kappa$ which is taken as 1. Below we refer to this algorithm as RTLS-SQP.

\begin{figure}[!ht]
    \centering
    \begin{minipage}[b]{0.24\textwidth}
        \centering
        \includegraphics[width=\linewidth]{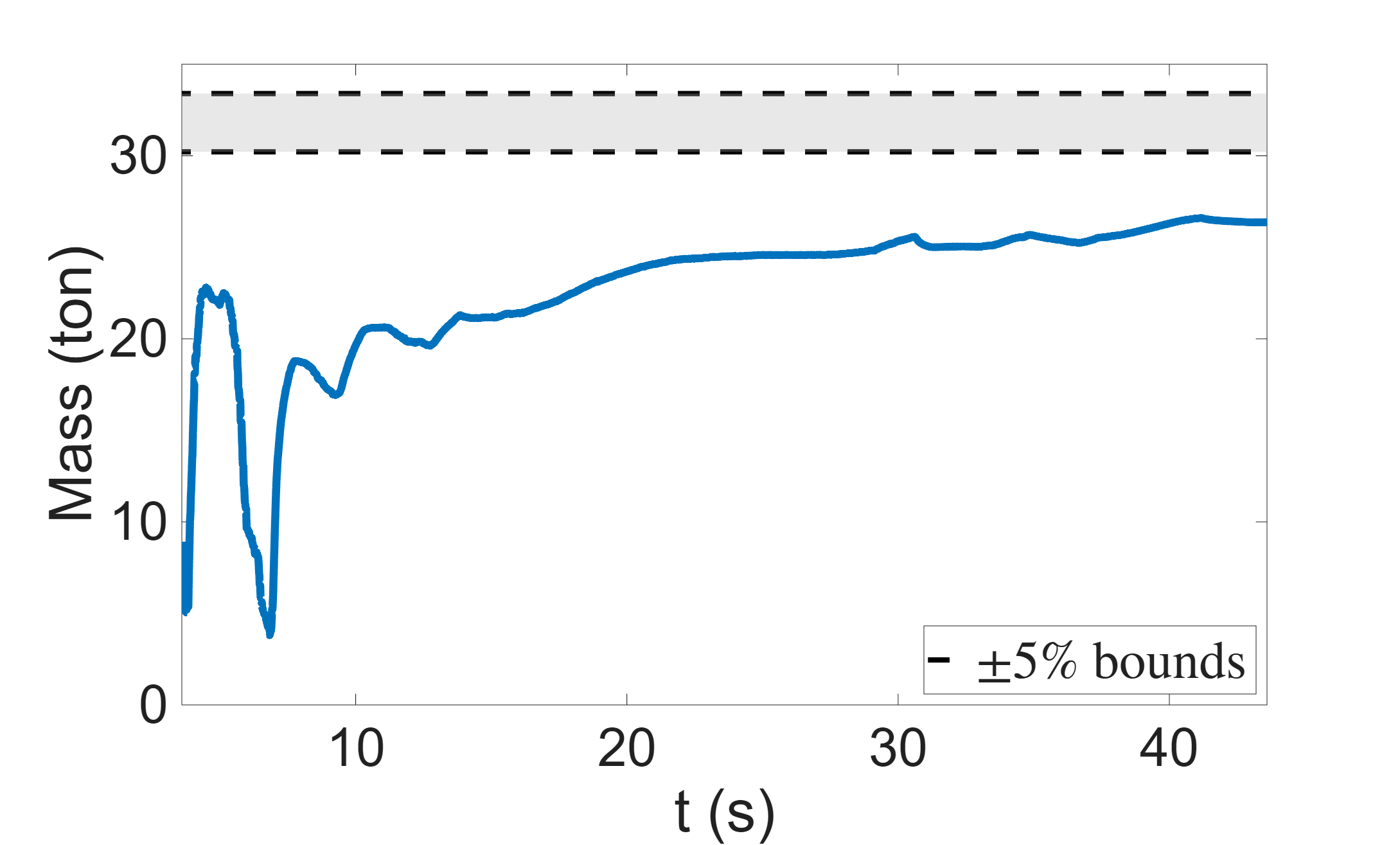}
        \caption*{(a) Mass estimate using RTLS-SQP on data from normal driving.}
    \end{minipage}\hfill
    \begin{minipage}[b]{0.24\textwidth}
        \centering
        \includegraphics[width=\linewidth]{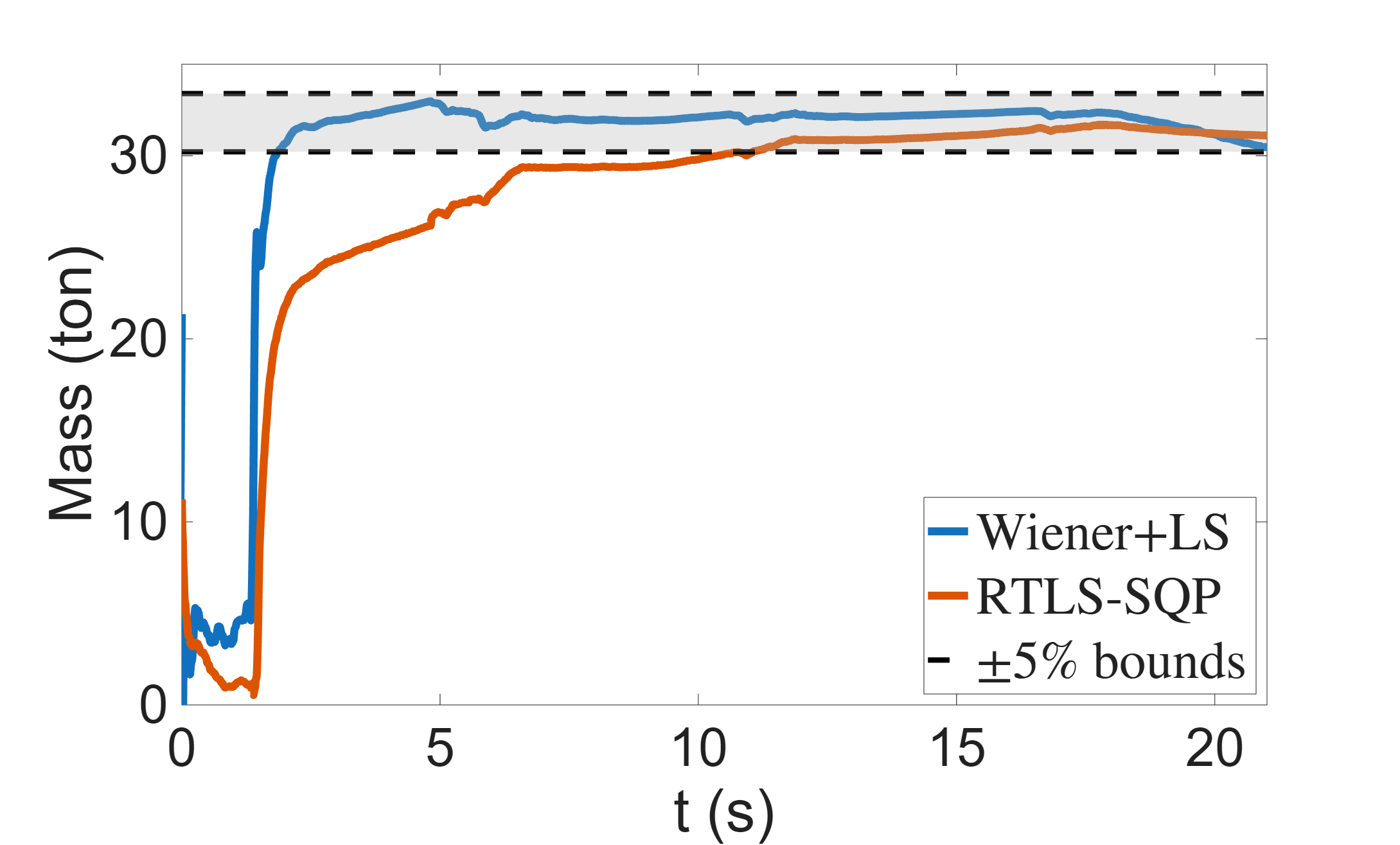}
        \caption*{(b) Mass estimate for RTLS-SQP and LS (with Wiener filtered acceleration).}
    \end{minipage}
    \caption{Estimation results for RTLS-SQP.}
    \label{fig:rtls}
    
\end{figure}

Fig. \ref{fig:rtls}(a) shows the resulting estimate on data from normal driving conditions for the truck having mass 31.8 tons. The velocity profile is similar to the ones shown in Fig. \ref{250625figure30}. In comparison Fig. \ref{fig:rtls}(b) shows the result when one of the trajectories optimized for minimum distance shown in Figs. \ref{fig:design_results_all}(e)-(h) are used as data. Clearly, the driving profile design has been beneficial for the estimation accuracy. This figure also shows the least-squares estimate using the Wiener filtered acceleration signal. Comparing with the RTLS-SQP estimate we conclude that using the Wiener filter method proposed in Section \ref{filteringsec} is competitive with respect to state-of-the art.   

\subsection{Summary}

From the comparisons between the results of the minimum-time and minimum-distance designs made above, we can conclude that the results are robust. Even with non-perfect following of the designed driving profile, the outcome is still in line with the design objective: a minimum-time design leads to shorter time used than a minimum-distance design, whereas a minimum-distance design leads to a shorter traveled distance than a minimum-time design, and both give much improved accuracy as compared to normal driving.

For the minimum-time design, we observe a clear bias in the estimates for both payload cases, whereas the bias is minor for the minimum-distance design. A potential reason for this could be that the minimum distance designs result in more high-frequent excitation that counter static non-linear phenomena in the power-train, like stiction.

We also remark that the difference in time and distance between the two designs can be substantial, with gains up to 35\% in our tests. In particular, when the road length is limited, a minimum-distance design may therefore be preferable. 

\section{Conclusions}
\label{sec:conclude}
This paper presented a framework for optimal driving profile design for real-world mass estimation in heavy-duty vehicles. The estimation problem is formulated under practical constraints on acceleration, velocity, and accuracy. Three representative objectives are considered: minimum-time, minimum experiment cost (minimum-distance), and maximum accuracy under fixed time. We also showed how nuisance parameters, i.e., parameters not of direct relevance to the mass, can be handled. Theoretical analysis provides insights into the optimal profiles, including feasibility conditions, key ratios between velocity and acceleration bounds, and trade-offs between time optimal and distance optimal solutions. The approach was validated through Scania truck experiments with two distinct payloads. Despite implementation challenges due to the simple control system, the designed inputs achieved estimation errors consistent with the theoretical ones. These findings confirm the method’s applicability in practice and its potential for accurate mass estimation. A separate finding is that the minimum-distance designs, which generally use lower velocities, yielded mass estimates with less bias than minimum-time designs. 

We also showed, using real-world data, that driving profile design also can give substantial improvements over normal driving profiles for recursive total-least squares. This study also indicates that Wiener filtering combined with least-squares may be competitive in comparison with RTLS.


An independent contribution was the development of a non-causal Wiener filter where the parameters are tuned using the Empirical Bayes method. This filter was used in the real-world tests to filter accelerometer signals. The advantage over standard causal low-pass filters commonly used for this type of filtering is that no phase-lag was introduced. Another advantage was that the filter parameters are tuned on the data to which the filter is to be applied. 

A natural next step in these developments is the implementation on an autonomous vehicle. With a more sophisticated control system, improved tracking of the designed profile can be achieved, leading to more reliable results.
Another interesting topic would be to investigate other experimental costs. For example, the idea of stealthy experiments \cite{Potters:14a} could be pursued to minimize deviations from normal driving behavior while improving the mass estimation accuracy.


\appendices

\section{Extension to refined longitudinal models}
\label{modelapp}

In this appendix we briefly comment on how to extend the proposed method to more elaborate models than \eqref{newton}. 

\subsection{Model refinements}
Detailed models for the components in \eqref{Fres}--\eqref{newton} can be found in Section 6.5 of \cite{Isermann:21}.  Returning to the decomposition \eqref{Fres} of the resulting force, the traction force $F_{\text{trac}}$ is due to the engine torque, accounting for drivetrain efficiency. The rolling distance can be modelled as

\begin{align}
\label{Froll}
    F_{\text{rolling}}=c_{\text{r}0}+c_{\text{r}1}v+c_{r4}v^4
\end{align}

\noindent where $v$ is the longitudinal velocity, and where $c_{\text{r}i}$, $i\in\{0,1,4\}$, are proportional to $mg$. The air resistance can be modelled as
\begin{align}
\label{Fair}
    F_{\text{air}}=c_{\text{air}}(v-v_{\text{wind}})^2
\end{align}
where $c_{\text{air}}=\frac{1}{2}\rho C_{\text{D}}A$, where in turn $\rho$ is the air density, $ C_{\text{D}}$ is the drag coefficient, $A$ is the frontal area, and $v_{\text{wind}}$ is the wind velocity. The gravitational force is given by

\begin{align}
\label{Fgravity}
    F_{\text{gravity}}=mg\sin \alpha
\end{align}

\noindent where $\alpha$ is the inclination of the road. Combining \eqref{Fres}, \eqref{Froll}-\eqref{Fgravity} gives the model
\begin{align}
\label{Ftrac}
    F_{\text{trac}}(k)=&ma(k)+c_{\text{r}0}+c_{\text{r}1}v(k)+c_{r4}v^4(k)\nonumber\\
    &+c_{\text{air}}(v(k)-v_{\text{wind}}(k))^2+mg\sin \alpha(k)
\end{align}
In, e.g.,  \cite{Koide:25} it is argued that the inertial measurement unit (IMU) approximately provides the signal
\begin{align}
\label{IMU}
    a_{\text{IMU}}(k)=a(k)+g\sin \alpha(k)
\end{align}
which inserted in \eqref{Ftrac} gives the model
\begin{align}
\label{Ftra2c}
    F_{\text{trac}}(k)=&m\, a_{\text{IMU}}(k)+c_{\text{r}0}+c_{\text{r}1}v(k)+c_{r4}v^4(k)\nonumber\\
    &+c_{\text{air}}(v(k)-v_{\text{wind}}(k))^2
\end{align}
The model used in \cite{Koide:25} is of this form, save for that the wind contribution is considered negligible in comparison to the vehicle velocity and that the traction force is computed from the torque and powertrain efficiency, giving
\begin{align}
\label{Ftrac2}
    F_{\text{trac}}(k)=&\varphi^T(k)\theta,\;
    \varphi(k)=\begin{bmatrix}
       a_{\text{IMU}}(k) & \hspace{-0.2cm} 1 & \hspace{-0.2cm}v(k) & \hspace{-0.2cm}v^4(k)&\hspace{-0.2cm} v^2(k)
    \end{bmatrix},\nonumber \\
    \theta=&\begin{bmatrix}
         m &c_{\text{r}0}&c_{\text{r}1}&c_{\text{r}4}&c_{\text{air}}
    \end{bmatrix}\nonumber
\end{align}
At low speed this model can be simplified. As noted experimentally in \cite{Rahkola:25} for a range of trucks, the rolling resistance for heavy duty trucks is almost independent of the velocity up to around 10 m/s, thus for velocities lower than 10 m/s, $F_{\text{rolling}}\approx c_{\text{r}0}$ with good accuracy. However, for the remaining part of this appendix we will assume that the model is given by \eqref{Ftrac2}.

\subsection{Least-squares estimation}
The least-squares estimate of $\theta$ in \eqref{Ftrac2} is given by
\begin{align}
    \hat{\theta}=&R^{-1}(N)\sum_{k=1}^N F_{\text{trac}}(k)\varphi(k),
    \, R(N)=\sum_{k=1}^N\varphi(k)\varphi^T(k) \nonumber
\end{align}
and assuming that additive Gaussian white noise $e(k)\sim{\mathcal N}(0,\sigma_e^2)$ contaminates $F_{\text{trac}}(k)$ (similar as for $F_{\text{res}}(k)$ in \eqref{newton3}) gives that \cite{Ljung:99}
\begin{align}
    \hat{\theta}\sim {\mathcal N}(\theta,P),\quad P=\sigma_e^2 R^{-1}(N)
\end{align}

\subsection{Driving profile design using an extended model}
\label{sec:nuis}
When the model is changed, the only quantity that changes in the optimal driving profile design problem  is the quality constraint \eqref{aomqc}.
Consider now that $m$ is the only element of $\theta$ of interest and that we use the quality measure \eqref{relerror}. 
Then \eqref{eq:J''} gives

\begin{align}
    J_{\text{app}}''(\theta_\circ)=\begin{bmatrix}
       \frac{2}{m_{\circ}^2}  & {\mathbf 0}_{1\times 4}\\
       {\mathbf 0}_{4\times 1} & {\mathbf 0_{4,4}}
    \end{bmatrix}\nonumber,
\end{align}
where ${\mathbf 0}_{m,n}$ is the $m\times n$ zero matrix. Inserting this in the quality constraint \eqref{aomqc} and noting that the only non-trivial constraint concerns the (1,1)-element, which can be written as \eqref{qualityconstr}, with $a(k)$ replaced by $a_{\text{IMU}}(k)$. This means that the optimal driving profile design problem remains the same as with the simpler model \eqref{newton}. The only difference is that the design gives the desired acceleration for the IMU, rather than the lateral acceleration of the vehicle.

\section{Rank constrained formulation}
\label{rankconstrainedapp}
 We follow the lifting approach taken in \cite{Manchester:10}, also pursued in, e.g., \cite{Larsson:15a}. It holds that 
 \begin{align*}
     U=uu^T\quad \Leftrightarrow \quad\begin{bmatrix}
         U & u\\u^T & 1
     \end{bmatrix}\succeq 0,\; \text{Rank}\; \begin{bmatrix}
         U & u\\u^T& 1
     \end{bmatrix}=1.
 \end{align*}
Introducing $A=aa^T$, we have 
\begin{align}
    A=aa^T=Fuu^TF^T=FUF^T,
        \label{eq:A}
\end{align}
from which it follows that the application quality constraint \eqref{constrainta} can be expressed as
\begin{align*}
    \text{Tr }UF^TF\geq \frac{2\sigma_e^2 \gamma\chi_\alpha^2(1)}{m_\circ^2}.
\end{align*}
It is easy to verify that the inequalities $a_{\min}\leq a(k)\leq a_{\max},$ occuring in \eqref{constraintc}
can be expressed as 
\begin{align*}
    a(k)^2-(a_{\min}+a_{\max})a(k)+a_{\min}a_{\max}\leq 0,
\end{align*}
which we can write as
\begin{align*}
    A(k,k)-(a_{\min}+a_{\max})a(k)+a_{\min}a_{\max}\leq 0.
\end{align*}
The equation above is a linear constraint in $U$ and $u$, since $A$ and $a$ are linear in $U$ and $u$, respectively (see \eqref{eq:A} and \eqref{a}). 

Velocity at time \(k\) is given by \eqref{velocity}, and the constraint \( v_{\min} \leq v_k \leq v_{\max} \) in \eqref{constraintd} can be expressed using the same technique,
\begin{align*}
T_s^2 \left( \sum_{l=1}^k a(\ell) \right)^2 
- (v_{\min} + v_{\max}) T_s \sum_{l=1}^k a(\ell) 
+ v_{\min} v_{\max} \leq 0,
\end{align*}
which, in lifted form, becomes:
\begin{align*}
T_s^2 \sum_{\ell_1=1}^{k} \sum_{\ell_2=1}^{k} A(\ell_1, \ell_2)
&- (v_{\min} + v_{\max}) T_s \sum_{\ell=1}^k a(\ell) \\
&+ v_{\min} v_{\max} \leq 0.
\end{align*}
This is a linear constraint with respect to $(A, a)$ and therefore, by way of \eqref{eq:A} and \eqref{a}, also with respect to $(U, u)$.

Since \eqref{constraintb} is linear in \( u \), all constraints in \eqref{constraints} can be expressed as linear in \( U \) and \( u \). Thus, the optimal input design problems in Section~\ref{oidsec}, based on \eqref{constraints}, can be formulated with \( U \) and \( u \) as decision variables, a linear objective, linear constraints, and a semi-definite constraint
\begin{align*}
    U\succeq 0,
\end{align*}
and finally, the non-convex constraint 
\begin{align}
\label{rankconstraint2}
    \text{Rank}\; \begin{bmatrix}
         U & u\\u^T& 1
     \end{bmatrix}=1.
\end{align}
Without \eqref{rankconstraint2}, the problem becomes convex. A common approach is to relax such a rank constraint using penalty functions that encourage \( U = uu^T \), as in \cite{Manchester:10,Larsson:15a}. One example is nuclear norm relaxation \cite{recht2010guaranteed}, where the nuclear norm \( \|A\|_* \), defined as the sum of the singular values of \( A \), serves as a convex surrogate for the rank function. Instead of enforcing \( \operatorname{Rank}\left[\begin{smallmatrix} U & u \\ u^T & 1 \end{smallmatrix}\right] = 1 \), we impose the convex constraint \( \|\left[\begin{smallmatrix} U & u \\ u^T & 1 \end{smallmatrix}\right]\|_* \leq \eta \) for some chosen threshold \( \eta \), which results in a convex SDP without introducing penalty terms into the objective.

There also exist solvers, e.g., LMIRANK \cite{Orsi:06}, that allow one to keep \eqref{rankconstrainedapp}, meaning that in case the solver finds a solution, it will be optimal for the original problem. 

\section{Derivation of the Velocity and Distance}
\label{sec:vel_dis}

\subsection{Velocity derivation}

Let \( T_s \) denote sampling time. Acceleration is assumed piecewise constant over each interval \( [(k-1)T_s,\, kT_s) \), i.e., the acceleration is a(k) for \( t \in [(k-1)T_s,\, kT_s) \).  
Under this assumption, velocity evolves linearly within each interval, i.e., the velocity at $s\in[(l-1)T_s,lT_s)$ is
\begin{align*}
   T_s \sum_{q=1}^{\ell-1} a(q) + (s - (\ell-1)T_s) a(\ell),
\end{align*}
where the first term gives the velocity at \( (\ell-1)T_s \), and the second term adds the linear contribution from constant acceleration \( a(\ell) \) within the interval. In particular, the velocity at time \( kT_s \) is given by:
\begin{align*}
    v(k) = T_s \sum_{\ell=1}^{k} a(\ell).
\end{align*}

\subsection{Distance derivation}

The distance at time \(  k T_s \) is given by:
\begin{align*}
    d(k) &=\sum_{\ell=1}^k \int_{(\ell-1)T_s}^{\ell T_s} \Bigl(T_s \sum_{q=1}^{\ell-1} a(q) + (s - (\ell-1)T_s) a(\ell)\Bigr)\, ds\\
   & =T_s^2\sum_{\ell=1}^{k}\frac{2(k-\ell)+1}{2}\; a(\ell).
\end{align*}

\section{Proof of Theorem 1}
\label{proof:t2}
\begin{proof}
By construction, \(n_+\) and \(n_-\) are positive integers. Moreover,
\[
    \bar a_+ = \frac{v_{\max}-v_{\min}}{T_s n_+} \leq a_{\max},
    \qquad
    |\bar a_-| = \frac{v_{\max}-v_{\min}}{T_s n_-} \leq |a_{\min}|,
\]
and hence the acceleration bounds are satisfied, i.e.,
\[
    a_{\min} \leq \bar a_- < 0 < \bar a_+ \leq a_{\max}.
\]

Each period consists of \( n_+ \) steps of \( \bar a_+ \) followed by \( n_- \) steps of \( \bar a_- \). By construction,
\[
    T_s n_+ \bar a_+ = v_{\max} - v_{\min},
    \qquad
    T_s n_- |\bar a_-| = v_{\max} - v_{\min}.
\]
Therefore, during the acceleration phase the velocity increases from \(v_{\min}\) to \(v_{\max}\), and during the deceleration phase it decreases from \(v_{\max}\) back to \(v_{\min}\). Hence, the velocity remains within \( [v_{\min},v_{\max}] \) throughout the horizon.

The total excitation energy over \( M \) periods is
\begin{align*}
    \sum_{k=1}^{N} a^2(k)
    =
    M\left(n_+\bar a_+^2+n_-\bar a_-^2\right).
\end{align*}
The accuracy constraint $\sum_{k=1}^N a^2(k) \geq R_{\mathrm{designed}}$ is satisfied if
\begin{align*}
    M \geq
    \frac{R_{\mathrm{designed}}}
    {n_+\bar a_+^2+n_-\bar a_-^2}.
\end{align*}
This completes the proof.
\end{proof}

\section{Proof of Theorem 2}
\label{sec:special}

\begin{proof}
Each cycle consists of acceleration from \( v_{\min} \) to \( v_1 \) with \( a_{\max} \) over time \( t_1 = \Delta v_1 / a_{\max} \), and deceleration back to \( v_{\min} \) with \( a_{\min} \) over time \( t_2 = \Delta v_1 / |a_{\min}| \), where \( \Delta v_1 = v_1 - v_{\min} \). The total duration of one cycle is
\[
p = t_1 + t_2 = \frac{(\left| a_{\min} \right| + a_{\max}) \Delta v_1}{\left| a_{\min} \right| a_{\max}}.
\]
The excitation energy per cycle is
\begin{align*}
    a_p =  \frac{1}{T_s} \left( t_1 a_{\max}^2 + t_2 |a_{\min}|^2 \right) = \frac{\Delta v_1 (|a_{\min}| + a_{\max})}{T_s}.
\end{align*}
The distance traveled in one cycle is
\begin{align*}
d_p &= p \cdot v_{\min} + p \cdot \frac{\Delta v_1}{2} \\
&= \frac{(|a_{\min}| + a_{\max}) \, \Delta v_1 \, (v_1 + v_{\min})}{2 |a_{\min}| a_{\max}}.
\end{align*}
The final acceleration from \( v_{\min} \) to \( v_{\max} \) takes time \( t_3 = (v_{\max} - v_{\min}) / a_{\max} \), with excitation energy and distance traveled as follows.
\begin{align*}
a_{\text{final}} &=\frac{t_3 a_{\max}^2}{T_s} =\frac{(v_{\max} - v_{\min}) a_{\max}}{T_s},\\
d_{\text{final}} &=\frac{t_3 (v_{\max} + v_{\min})}{2}=\frac{v_{\max}^2 - v_{\min}^2}{2a_{\max}} .
\end{align*}
The total excitation energy is the sum of that from all cycles and the final segment, i.e.,
\begin{align*}
R(N)
&= M \cdot a_p + a_{\text{final}} \\
&= \frac{1}{T_s} \Bigl( M  (|a_{\min}| + a_{\max})\Delta v_1 
+ (v_{\max} - v_{\min}) a_{\max} \Bigr).
\end{align*}
To have $R(N)= R_{\text{designed}}$, we need the number of cycles $M$ satisfying
\begin{align*}
M (|a_{\min}| + a_{\max})\Delta v_1  =
T_sR_{\text{designed}}+(v_{\min} - v_{\max}) a_{\max}.
\end{align*}
The total distance, as the sum of contributions from the periodic and final acceleration segments, simplifies in the limiting case \( v_1 \to v_{\min} \). Then,
\begin{align}
d_{\text{distance}} 
=& M  d_p +d_{\text{final}}  \nonumber\\
=& \frac{M(|a_{\min}| + a_{\max}) \Delta v_1 v_{\min}}{ |a_{\min}| a_{\max}}+\frac{v_{\max}^2 - v_{\min}^2}{2 a_{\max}}  \nonumber\\
=&(T_sR_{\text{designed}} + v_{\min} a_{\max} -v_{\max}a_{\max} ) 
    \frac{v_{\min}}{|a_{\min}| a_{\max}} \nonumber\\ &+ \frac{v_{\max}^2 - v_{\min}^2}{2 a_{\max}} \nonumber\\
=&\frac{T_sR_{\text{designed}} v_{\min}}{|a_{\min}| a_{\max}}  - \frac{(v_{\max} - v_{\min}) v_{\min}}{|a_{\min}|} + \frac{v_{\max}^2 - v_{\min}^2}{2 a_{\max}}.
    \label{eq:ddistance}
\end{align}
To minimize the total distance with respect to \( v_{\max} \), we take the derivative
\[
\frac{\partial d_{\text{distance}}}{\partial v_{\max}} 
= - \frac{v_{\min}}{|a_{\min}|} + \frac{v_{\max}}{a_{\max}}.
\]
Setting the derivative to zero yields the optimality condition
\begin{align}
    \label{eq:amax}
    \frac{v_{\max}}{v_{\min}} = \frac{a_{\max}}{|a_{\min}|}.
\end{align}
By using \eqref{eq:amax}, then \eqref{eq:ddistance} comes to
\begin{align*}
d^\star = \frac{T_sR_{\text{designed}} v_{\min}}{|a_{\min}| a_{\max}}
- \frac{(|a_{\min}|-a_{\max})^2v_{\min}^2}{2|a_{\min}|^2 a_{\max}}.
\end{align*}
This completes the proof.
\end{proof}

\section{Proof of Theorem 3}

\label{theorem3}
\begin{proof}

In the minimum-time solution with periodic acceleration, the vehicle repeatedly accelerates from \(v_{\min}\) to \(v_{\max}\) using \(a_{\max}>0\), and then decelerates back to \(v_{\min}\) using \(a_{\min}<0\). Each cycle thus consists of two symmetric phases, i.e.,

\begin{itemize}
  \item \(d_+\): distance during the acceleration phase \((v_{\min}\!\to v_{\max})\);
  \item \(d_-\): distance during the deceleration phase \((v_{\max}\!\to v_{\min})\);
  \item \(d_{\mathrm{cyc}}\!= d_+ + d_-\): total distance per cycle.
\end{itemize}

\paragraph*{Distance in a single cycle}  
Using the kinematic relation, the distance for a full acceleration–deceleration cycle is
\begin{align*}
    d_{\mathrm{cyc}} &= d_+ + d_- 
= \frac{v_{\max}^2 - v_{\min}^2}{2a_{\max}} 
  + \frac{v_{\max}^2 - v_{\min}^2}{2|a_{\min}|} \\
&= \frac{v_{\max}^2 - v_{\min}^2}{2}
    \left(\frac{1}{a_{\max}} + \frac{1}{|a_{\min}|} \right).
\end{align*}

\textit{Distance of time minimization objective:}  
By Theorem~1, the minimum number of cycles required to satisfy the designed excitation-energy \(R_{\mathrm{designed}}\) is $M \;=\;
\frac{R_{\mathrm{designed}}}{n_+ a_{\max}^2 + n_- a_{\min}^2}.$ Then,
\begin{align}
    d_{\text{time}} &= M\,d_{\mathrm{cyc}} = R_{\mathrm{designed}}T_s
\,\frac{v_{\max}+v_{\min}}{2a_{\max}|a_{\min}|}\nonumber\\
&= \frac{R_{\mathrm{designed}}T_s}{a_{\max}|a_{\min}|}
    \left(v_{\min} + \tfrac{\Delta v}{2}\right),
\label{eq:dtime-final}
\end{align}
where \( \Delta v = v_{\max} - v_{\min} \).

\textit{Distance gap:}  
The difference between \eqref{eq:dtime-final} and \eqref{eq:ddistance} gives the distance gap,
\begin{align}
    \Delta d(\Delta v)
    = &d_{\text{time}}-d_{\text{distance}}\\
    =& -\,\frac{\Delta v^{2}}{2a_{\max}}
      + \frac{\Delta v}{|a_{\min}|}
        \Bigl(
           v_{\min} + \frac{T_sR_{\mathrm{designed}}}{2a_{\max}}\\
          & -\frac{v_{\min} |a_{\min}|}{a_{\max}}
        \Bigr).
        \label{eq:deltad}
\end{align}
Equation~\eqref{eq:deltad} is a downward-opening quadratic in \(\Delta v\).  
Its axis of symmetry is located at
\[
\Delta v^{\star}
   = \frac{a_{\max}-|a_{\min}|}{|a_{\min}|}\,v_{\min}
     + \frac{T_sR_{\mathrm{designed}}}{2|a_{\min}|}.
\]

\textit{Monotonicity to the left of the axis:}  
From Theorem~2 the admissible velocity upper bound satisfies 
\(v_{\max}= (a_{\max}/|a_{\min}|)\,v_{\min}\).  
Consequently
\(
\Delta v = v_{\max}-v_{\min} \le \Delta v^{\star},
\)
so the operating point is located on the left‐hand side of the symmetry axis of the downward-opening quadratic \(\Delta d(\Delta v)\).  
Because the derivative is positive in this interval, \(\Delta d(\Delta v)\) is strictly increasing.  Hence, a larger admissible velocity range \(\Delta v\) always produces a larger travel distance gap \(\Delta d\).
\end{proof}

\bibliography{ref}

\begin{thebibliography}{10}
\providecommand{\url}[1]{#1}
\csname url@samestyle\endcsname
\providecommand{\newblock}{\relax}
\providecommand{\bibinfo}[2]{#2}
\providecommand{\BIBentrySTDinterwordspacing}{\spaceskip=0pt\relax}
\providecommand{\BIBentryALTinterwordstretchfactor}{4}
\providecommand{\BIBentryALTinterwordspacing}{\spaceskip=\fontdimen2\font plus
\BIBentryALTinterwordstretchfactor\fontdimen3\font minus \fontdimen4\font\relax}
\providecommand{\BIBforeignlanguage}[2]{{%
\expandafter\ifx\csname l@#1\endcsname\relax
\typeout{** WARNING: IEEEtran.bst: No hyphenation pattern has been}%
\typeout{** loaded for the language `#1'. Using the pattern for}%
\typeout{** the default language instead.}%
\else
\language=\csname l@#1\endcsname
\fi
#2}}
\providecommand{\BIBdecl}{\relax}
\BIBdecl

\bibitem{li2017two}
B.~Li, J.~Zhang, H.~Du, and W.~Li, ``Two-layer structure based adaptive estimation for vehicle mass and road slope under longitudinal motion,'' \emph{Measurement}, vol.~95, pp. 439--455, 2017.

\bibitem{mcintyre2009two}
M.~L. McIntyre, T.~J. Ghotikar, A.~Vahidi, X.~Song, and D.~M. Dawson, ``A two-stage lyapunov-based estimator for estimation of vehicle mass and road grade,'' \emph{IEEE Transactions on Vehicular Technology}, vol.~58, no.~7, pp. 3177--3185, 2009.

\bibitem{pence2009sprung}
B.~L. Pence, H.~K. Fathy, and J.~L. Stein, ``Sprung mass estimation for off-road vehicles via base-excitation suspension dynamics and recursive least squares,'' in \emph{Proceedings of American Control Conference}, 2009, pp. 5043--5048.

\bibitem{yu2022mass}
Z.~Yu, X.~Hou, B.~Leng, and Y.~Huang, ``Mass estimation method for intelligent vehicles based on fusion of machine learning and vehicle dynamic model,'' \emph{Autonomous Intelligent Systems}, vol.~2, no.~1, p.~4, 2022.

\bibitem{fathy2008online}
H.~K. Fathy, D.~Kang, and J.~L. Stein, ``Online vehicle mass estimation using recursive least squares and supervisory data extraction,'' in \emph{Proceedings of American Control Conference}, 2008, pp. 1842--1848.

\bibitem{zarringhalam2012comparative}
R.~Zarringhalam, A.~Rezaeian, W.~Melek, A.~Khajepour, S.-k. Chen, and N.~Moshchuk, ``A comparative study on identification of vehicle inertial parameters,'' in \emph{Proceedings of American Control Conference}, 2012, pp. 3599--3604.

\bibitem{lundin2012estimation}
B.~Lundin and A.~Olsson, ``Estimation of vehicle mass using an extended {K}alman filter,'' Master thesis, Chalmers University of Technology, Gothenburg, Sweden, 2012.

\bibitem{lingman2002road}
P.~Lingman and B.~Schmidtbauer, ``Road slope and vehicle mass estimation using {K}alman filtering,'' \emph{Vehicle System Dynamics}, vol.~37, no. sup1, pp. 12--23, 2002.

\bibitem{winstead2005estimation}
V.~Winstead and I.~V. Kolmanovsky, ``Estimation of road grade and vehicle mass via model predictive control,'' in \emph{Proceedings of IEEE Conference on Control Applications}, 2005, pp. 1588--1593.

\bibitem{zhang2024identification}
X.~Zhang, J.~He, X.~Hua, and Z.~Chen, ``Identification of time-varying stiffness with unknown mass distribution based on extended {K}alman filter,'' \emph{Mechanical Systems and Signal Processing}, vol. 211, p. 111218, 2024.

\bibitem{huang2014real}
X.~Huang and J.~Wang, ``Real-time estimation of center of gravity position for lightweight vehicles using combined {AKF--EKF} method,'' \emph{IEEE Transactions on Vehicular Technology}, vol.~63, no.~9, pp. 4221--4231, 2014.

\bibitem{altmannshofer2016robust}
S.~Altmannshofer and C.~Endisch, ``Robust vehicle mass and driving resistance estimation,'' in \emph{Proceedings of American Control Conference}, 2016, pp. 6869--6874.

\bibitem{hong2014novel}
S.~Hong, C.~Lee, F.~Borrelli, and J.~K. Hedrick, ``A novel approach for vehicle inertial parameter identification using a dual {K}alman filter,'' \emph{IEEE Transactions on Intelligent Transportation Systems}, vol.~16, no.~1, pp. 151--161, 2014.

\bibitem{boada2019sensor}
B.~L. Boada, M.~J.~L. Boada, and H.~Zhang, ``Sensor fusion based on a dual {K}alman filter for estimation of road irregularities and vehicle mass under static and dynamic conditions,'' \emph{IEEE/ASME Transactions on Mechatronics}, vol.~24, no.~3, pp. 1075--1086, 2019.

\bibitem{rhode2012vehicle}
S.~Rhode and F.~Gauterin, ``Vehicle mass estimation using a total least-squares approach,'' in \emph{Proceedings of International IEEE Conference on Intelligent Transportation Systems}, 2012, pp. 1584--1589.

\bibitem{lin2018method}
N.~Lin, C.~Zong, and S.~Shi, ``The method of mass estimation considering system error in vehicle longitudinal dynamics,'' \emph{Energies}, vol.~12, no.~1, p.~52, 2018.

\bibitem{chor2023robust}
W.~T. Chor, C.~P. Tan, A.~Bakibillah, Z.~Pu, and J.~Y. Loo, ``Robust vehicle mass estimation using recursive least m-squares algorithm for intelligent vehicles,'' \emph{IEEE Transactions on Intelligent Vehicles}, vol.~9, no.~1, pp. 165--177, 2023.

\bibitem{vahidi2005recursive}
A.~Vahidi, A.~Stefanopoulou, and H.~Peng, ``Recursive least squares with forgetting for online estimation of vehicle mass and road grade: theory and experiments,'' \emph{Vehicle System Dynamics}, vol.~43, no.~1, pp. 31--55, 2005.

\bibitem{Koide:25}
H.~Koide, J.~Vayssettes, and G.~Mercère, ``Recursive total least squares with improved parameter tracking: Application to model-based vehicle mass estimation,'' \emph{Control Engineering Practice}, vol. 164, p. 106429, 2025.

\bibitem{pence2013recursive}
B.~L. Pence, H.~K. Fathy, and J.~L. Stein, ``Recursive estimation for reduced-order state-space models using polynomial chaos theory applied to vehicle mass estimation,'' \emph{IEEE Transactions on Control Systems Technology}, vol.~22, no.~1, pp. 224--229, 2013.

\bibitem{isbitirici2025data}
A.~Isbitirici, ``Data-driven mass estimation of heavy-duty vehicles,'' Ph.D. dissertation, University of Bologna, Bologna, Italy, 2025.

\bibitem{chen2023regenerative}
Z.~Chen, R.~Xiong, X.~Cai, Z.~Wang, and R.~Yang, ``Regenerative braking control strategy for distributed drive electric vehicles based on slope and mass co-estimation,'' \emph{IEEE Transactions on Intelligent Transportation Systems}, vol.~24, no.~12, pp. 14\,610--14\,619, 2023.

\bibitem{zhou2025hybrid}
Z.~Zhou, Y.~Wang, X.~Liu, Z.~Li, M.~Wu, and G.~Zhou, ``Hybrid of neural network and physics-based estimator for vehicle longitudinal dynamics modeling using limited driving data,'' \emph{IEEE Transactions on Intelligent Transportation Systems}, vol.~26, no.~10, pp. 16\,735--16\,746, 2025.

\bibitem{bevly2006integrating}
D.~M. Bevly, J.~Ryu, and J.~C. Gerdes, ``Integrating {INS} sensors with {GPS} measurements for continuous estimation of vehicle sideslip, roll, and tire cornering stiffness,'' \emph{IEEE Transactions on Intelligent Transportation Systems}, vol.~7, no.~4, pp. 483--493, 2006.

\bibitem{antunes2019implementation}
A.~Antunes, P.~Outeiro, C.~Cardeira, and P.~Oliveira, ``Implementation and testing of a sideslip estimation for a formula student prototype,'' \emph{Robotics and Autonomous Systems}, vol. 115, pp. 83--89, 2019.

\bibitem{cordeiro2019estimation}
R.~A. Cordeiro, A.~C. Victorino, J.~R. Azinheira, P.~A. Ferreira, E.~C. De~Paiva, and S.~S. Bueno, ``Estimation of vertical, lateral, and longitudinal tire forces in four-wheel vehicles using a delayed interconnected cascade-observer structure,'' \emph{IEEE/ASME Transactions on Mechatronics}, vol.~24, no.~2, pp. 561--571, 2019.

\bibitem{pham2019design}
T.-P. Pham, O.~Sename, and L.~Dugard, ``Design and experimental validation of an {$H_\infty$} observer for vehicle damper force estimation,'' \emph{IFAC-PapersOnLine}, vol.~52, no.~5, pp. 673--678, 2019.

\bibitem{pham2019real}
T.-P. Pham, O.~Sename, and L.~Dugard, ``Real-time damper force estimation of vehicle electrorheological suspension: A nonlinear parameter varying approach,'' \emph{IFAC-PapersOnLine}, vol.~52, no.~28, pp. 94--99, 2019.

\bibitem{pham2019unified}
T.-P. Pham, O.~Sename, and L.~Dugard, ``Unified {$\mathcal{H}_\infty$} observer for a class of nonlinear {Lipschitz} systems: Application to a real {ER} automotive suspension,'' \emph{IEEE Control Systems Letters}, vol.~3, no.~4, pp. 817--822, 2019.

\bibitem{RIBEIRO2021104924}
A.~Ribeiro, M.~Koyama, A.~Moutinho, E.~C. de~Paiva, and A.~Fioravanti, ``A comprehensive experimental validation of a scaled car-like vehicle: Lateral dynamics identification, stability analysis, and control application,'' \emph{Control Engineering Practice}, vol. 116, p. 104924, 2021.

\bibitem{Ljung:99}
L.~Ljung, \emph{System Identification: Theory for the User}, 2nd~ed.\hskip 1em plus 0.5em minus 0.4em\relax Englewood Cliffs, NJ: Prentice-Hall, 1999.

\bibitem{honorio2018persistently}
L.~M. Hon{\'o}rio, E.~B. Costa, E.~J. Oliveira, D.~de~Almeida~Fernandes, and A.~P.~G. Moreira, ``Persistently-exciting signal generation for optimal parameter estimation of constrained nonlinear dynamical systems,'' \emph{ISA Transactions}, vol.~77, pp. 231--241, 2018.

\bibitem{hjalmarsson2009system}
H.~Hjalmarsson, ``System identification of complex and structured systems,'' \emph{European Journal of Control}, vol.~15, no. 3--4, pp. 275--310, 2009.

\bibitem{annergren2017application}
M.~Annergren, C.~A. Larsson, H.~Hjalmarsson, X.~Bombois, and B.~Wahlberg, ``Application-oriented input design in system identification: Optimal input design for control,'' \emph{IEEE Control Systems Magazine}, vol.~37, no.~2, pp. 31--56, 2017.

\bibitem{Mehra:74}
R.~Mehra, ``Optimal input signals for parameter estimation in dynamic systems -- {S}urvey and new results,'' \emph{IEEE Transactions on Automatic Control}, vol.~19, no.~6, pp. 753--768, 1974.

\bibitem{Goodwin&Payne:77}
G.~C. Goodwin and R.~L. Payne, \emph{Dynamic System Identification: Experiment Design and Data Analysis}.\hskip 1em plus 0.5em minus 0.4em\relax New York: Academic Press, 1977.

\bibitem{Gevers&Ljung:86}
M.~Gevers and L.~Ljung, ``Optimal experiment designs with respect to the intended model application,'' \emph{Automatica}, vol.~22, no.~5, pp. 543--554, 1986.

\bibitem{Forssell&Ljung:00}
U.~Forssell and L.~Ljung, ``Some results on optimal experiment design,'' \emph{Automatica}, vol.~36, no.~5, pp. 749--756, 2000.

\bibitem{Atkinson&Bailey:01}
A.~Atkinson and R.~Bailey, ``One hundred years of the design of experiments on and off the pages of {B}iometrika,'' \emph{Biometrika}, vol.~88, pp. 53--97, 2001.

\bibitem{Gevers:11}
M.~Gevers, X.~Bombois, R.~Hildebrand, and G.~Solari, ``Optimal experiment design for open and closed-loop system identification,'' \emph{Communications in Information and Systems}, vol.~11, pp. 197--224, 2011.

\bibitem{Jansson:04a}
H.~Jansson and H.~Hjalmarsson, ``Input design via {LMI}s admitting frequency-wise model specifications in confidence regions,'' \emph{IEEE Transactions on Automatic Control}, vol.~50, no.~10, pp. 1534--1549, 2005.

\bibitem{Valenzuela:15a}
P.~Valenzuela, C.~Rojas, and H.~Hjalmarsson, ``A graph theoretical approach to input design for identification of nonlinear dynamical models,'' \emph{Automatica}, vol.~51, no.~1, pp. 233--242, 2015.

\bibitem{pang2016data}
Z.-H. Pang, G.-P. Liu, D.~Zhou, and D.~Sun, ``Data-driven control with input design-based data dropout compensation for networked nonlinear systems,'' \emph{IEEE Transactions on Control Systems Technology}, vol.~25, no.~2, pp. 628--636, 2016.

\bibitem{bombois2006least}
X.~Bombois, G.~Scorletti, M.~Gevers, P.~M. Van~den Hof, and R.~Hildebrand, ``Least costly identification experiment for control,'' \emph{Automatica}, vol.~42, no.~10, pp. 1651--1662, 2006.

\bibitem{larsson2014application}
C.~A. Larsson, ``Application-oriented experiment design for industrial model predictive control,'' Ph.D. dissertation, KTH Royal Institute of Technology, Stockholm, Sweden, 2014.

\bibitem{ebadat2017application}
A.~Ebadat, D.~Varagnolo, G.~Bottegal, B.~Wahlberg, and K.~H. Johansson, ``Application-oriented input design for room occupancy estimation algorithms,'' in \emph{Proceedings of IEEE Conference on Decision and Control}.\hskip 1em plus 0.5em minus 0.4em\relax IEEE, 2017, pp. 3417--3424.

\bibitem{Hjalmarsson:04a}
H.~Hjalmarsson, ``From experiment design to closed loop control,'' \emph{Automatica}, vol.~41, no.~3, pp. 393--438, 2005.

\bibitem{Annergren:12a}
M.~Annergren and C.~Larsson, ``{MOOSE:} {A} model based optimal input design toolbox,'' in \emph{Proceedings of IFAC Symposium on System Identification}, 2012, pp. 1535--1540.

\bibitem{Sigurdsson:24a}
G.~Sigurdsson, A.~Isaksson, M.~Lundh, H.~Hjalmarsson, and S.~Munusamy, ``Optimal experiment design for multivariable system identification using simultaneous excitation,'' in \emph{Proceedings of IFAC Symposium on System Identification}, 2024, pp. 544--549.

\bibitem{Rivera:09a}
D.~Rivera, H.~Lee, H.~Mittelmann, and M.~Braun, ``Constrained multisine input signals for plant-friendly identification of chemical process systems,'' \emph{Journal of Process Control}, vol.~19, no.~4, pp. 623--635, 2009.

\bibitem{Lundh:24}
M.~Lundh, S.~Munusamy, A.~Isaksson, H.~Hjalmarsson, and V.~Pinnamaraju, ``Optimal design of sequential excitation for identification of multi-variable systems,'' in \emph{Proceedings of IFAC Symposium on Advanced Control of Chemical Processes}, 2024, pp. 409--415.

\bibitem{soderstrom2018errors}
T.~S{\"o}derstr{\"o}m, \emph{Errors-in-Variables Methods in System Identification}.\hskip 1em plus 0.5em minus 0.4em\relax Springer, 2018.

\bibitem{Huffel:87}
S.~V. Huffel, ``Analysis of the total least squares problem and its use in parameter estimation,'' Ph.D. dissertation, Katholieke Universiteit Leuven, 1987.

\bibitem{Rhode:12}
S.~Rhode and F.~Gauterin, ``Vehicle mass estimation using a total least-squares approach,'' in \emph{2012 15th International IEEE Conference on Intelligent Transportation Systems}, 2012, pp. 1584--1589.

\bibitem{Rhode:13}
S.~Rhode and F.~Gauterin, ``Online estimation of vehicle driving resistance parameters with recursive least squares and recursive total least squares,'' in \emph{2013 IEEE Intelligent Vehicles Symposium (IV)}, 2013, pp. 269--276.

\bibitem{jessica2025}
J.~Ye, ``Application-oriented input design for mass estimation of heavy-duty autonomous mining trucks,'' Master thesis, KTH Royal Institute of Technology, Stockholm, Sweden, 2025.

\bibitem{Kailath:2000}
T.~Kailath, A.~H. Sayed, and B.~Hassibi, \emph{Linear estimation}.\hskip 1em plus 0.5em minus 0.4em\relax Prentice Hall, 2000.

\bibitem{Lehmann:98}
E.~L. Lehmann and G.~Casella, \emph{Theory of Point Estimation}, 2nd~ed.\hskip 1em plus 0.5em minus 0.4em\relax New York: John Wiley \& Sons, 1998.

\bibitem{RasmussenW:2006}
C.~E. Rasmussen and C.~K.~I. Williams, \emph{Gaussian Processes for Machine Learning}.\hskip 1em plus 0.5em minus 0.4em\relax MIT Press, 2006.

\bibitem{PillonettoCD:2011}
G.~Pillonetto, A.~Chiuso, and G.~De~Nicolao, ``Prediction error identification of linear systems: A nonparametric {G}aussian regression approach,'' \emph{Automatica}, vol.~47, no.~2, pp. 291--305, 2011.

\bibitem{Orsi:06}
R.~Orsi, U.~Helmke, and J.~B. Moore., ``A {N}ewton-like method for solving rank constrained linear matrix inequalities,'' \emph{Automatica}, vol.~42, no.~11, pp. 1875--1882, 2006.

\bibitem{lofberg2004yalmip}
J.~Lofberg, ``{YALMIP}: A toolbox for modeling and optimization in {MATLAB},'' in \emph{Proceedings of IEEE International Conference on Robotics and Automation}, 2004, pp. 284--289.

\bibitem{Gerencser:07}
L.~Gerencs\'{e}r, H.~Hjalmarsson, and J.~M{\aa}rtensson, ``Identification of {ARX} systems with non-stationary inputs - asymptotic analysis with application to adaptive input design,'' \emph{Automatica}, vol.~45, no.~3, pp. 623--633, 2009.

\bibitem{Huang:14b}
L.~Gerencs\'{e}r, H.~Hjalmarsson, and L.~Huang, ``Adaptive input design for {LTI} systems,'' \emph{IEEE Transactions on Automatic Control}, vol.~62, no.~5, pp. 2390--2405, 2016.

\bibitem{Koide:26}
H.~Koide, J.~Vayssettes, and G.~Mercère, ``A recursive parameter identification algorithm for nonlinear errors-in-variables models,'' \emph{IFAC Journal of Systems and Control}, vol.~35, p. 100381, 2026.

\bibitem{Koidethesis:26}
H.~Koide, ``Errors-in-variables model identification with application to online vehicle mass estimation,'' Doctoral thesis, Universit\'{e} De Poitiers, Poitiers, France, 2026.

\bibitem{Potters:14a}
M.~Potters, X.~Bombois, M.~Forgione, P.~Mod\'{e}n, M.~Lundh, H.~Hjalmarsson, and P.~{Van den Hof}, ``Optimal experiment design in closed loop with unknown nonlinear or implicit controllers using stealth identification,'' in \emph{Proceedings of European Control Conference}, 2014.

\bibitem{Isermann:21}
R.~Isermann, \emph{Automotive Control. Modeling and Control of Vehicles}.\hskip 1em plus 0.5em minus 0.4em\relax Springer Berlin, Heidelberg, 2021.

\bibitem{Rahkola:25}
P.~Rahkola, \emph{\BIBforeignlanguage{English}{Driving Resistance Measurements of Heavy-Duty Vehicles}}, ser. VTT Research Report No. VTT-R-00316-25.\hskip 1em plus 0.5em minus 0.4em\relax VTT Technical Research Centre of Finland, 2025.

\bibitem{Manchester:10}
I.~Manchester, ``Input design for system identification via convex relaxation,'' in \emph{Proceedings of IEEE Conference on Decision and Control}, 2010, pp. 2041--2046.

\bibitem{Larsson:15a}
C.~Larsson, C.~Rojas, X.~Bombois, and H.~Hjalmarsson, ``Experimental evaluation of model predictive control with excitation ({MPC-X}) on an industrial depropanizer,'' \emph{Journal of Process Control}, vol.~31, pp. 1--16, 2015.

\bibitem{recht2010guaranteed}
B.~Recht, M.~Fazel, and P.~A. Parrilo, ``Guaranteed minimum-rank solutions of linear matrix equations via nuclear norm minimization,'' \emph{SIAM Review}, vol.~52, no.~3, pp. 471--501, 2010.

\end{thebibliography}
\bibliographystyle{IEEEtran}


\begin{IEEEbiography}[{\raisebox{0pt}[1.25in][0pt]{\includegraphics[width=1in,height=1.25in,clip]{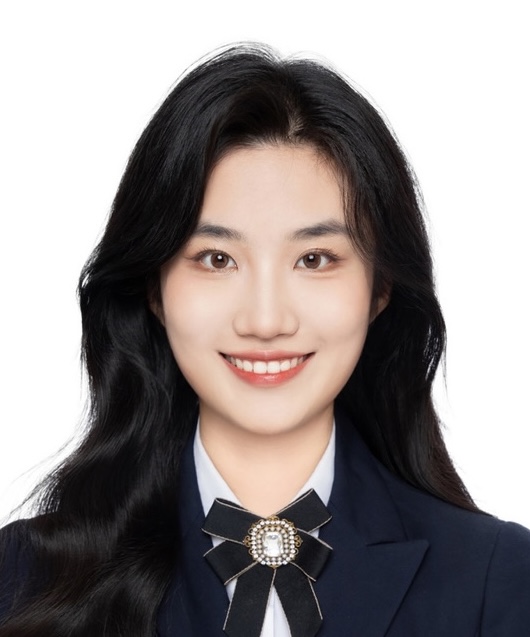}}}]{Le Wang} is currently pursuing a joint Ph.D. degree with KTH Royal Institute of Technology and Shanghai Jiao Tong University, majoring in electrical engineering and control science and engineering, respectively. She received her Bachelor's degree in Automation from Shandong University in 2019. Her research interests include system identification, input design, optimization algorithms, signal processing, and their related applications.
\end{IEEEbiography}


\begin{IEEEbiography}[{\raisebox{0pt}[1.25in][0pt]{\includegraphics[width=1in,height=1.25in,clip]{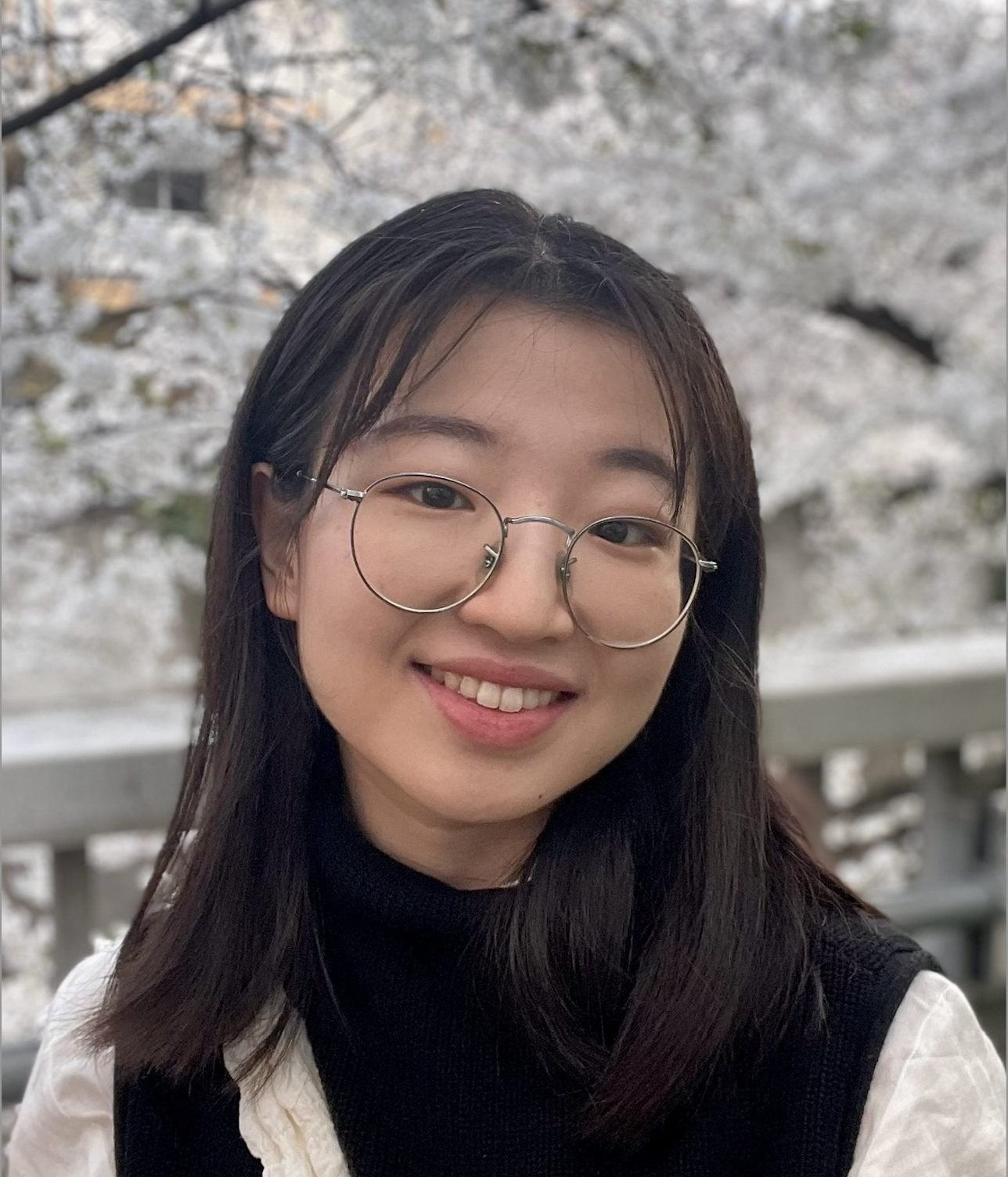}}}]{Jessica Ye} is currently pursuing a Master’s degree in Systems, Control and Robotics at KTH Royal Institute of Technology, where she also completed her Bachelor's degree in Electrical Engineering. In the spring of 2025, she conducted her thesis project on enhancing mass estimation through an application-oriented input design approach.
\end{IEEEbiography}


\begin{IEEEbiography}[
{\raisebox{0pt}[1.25in][0pt]{\includegraphics[width=1in,height=1.2in,clip]{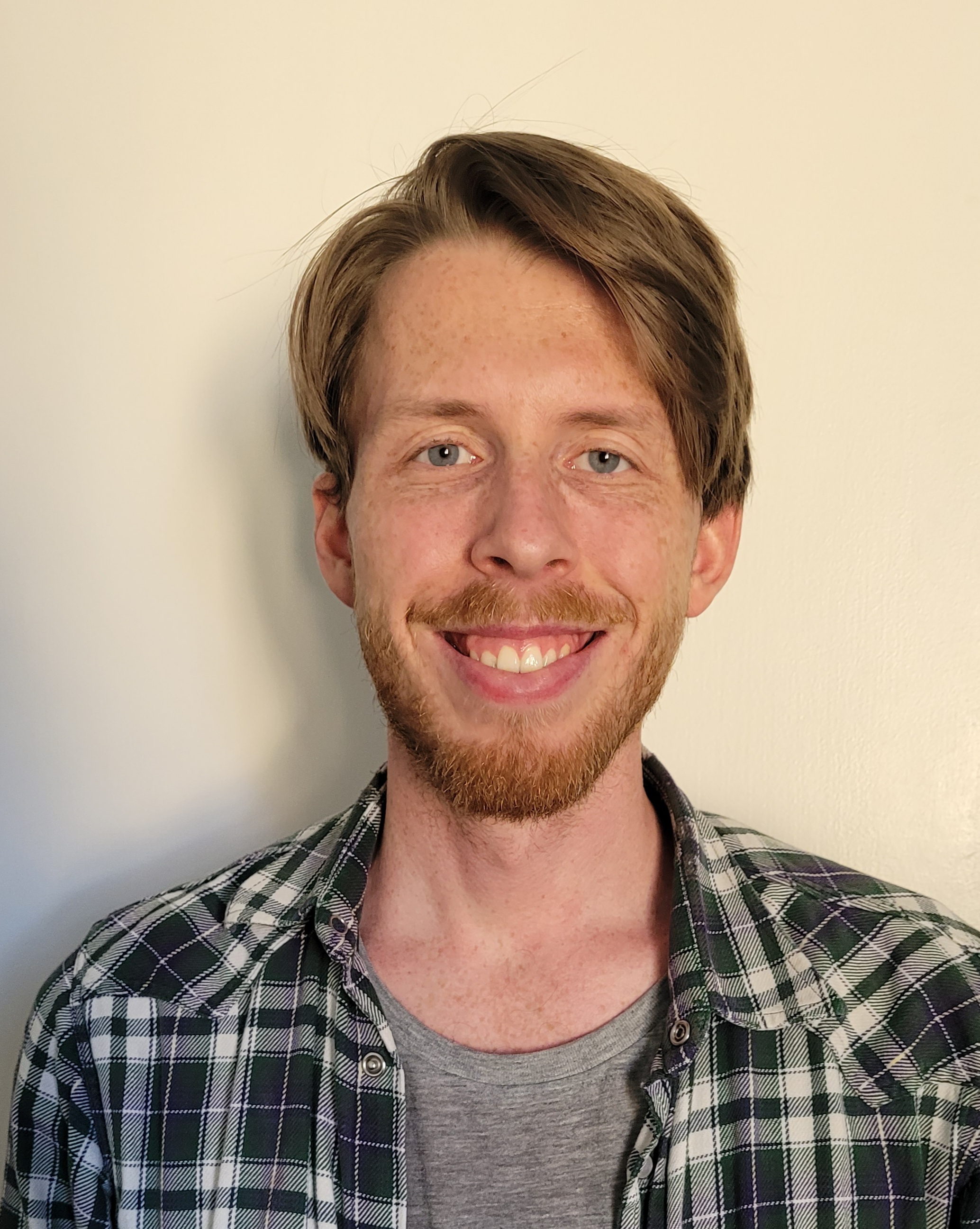}}}
]{Michael Refors} is an Embedded Software Developer at TRATON Group R\&D, specializing in powertrain control systems. His work focuses on the development of parameter estimation functions and energy consumption algorithms for advanced vehicle powertrains. Prior to his current role, Michael held embedded software development and architecture positions at Scania CV AB and Atlas Copco Industrial Technique AB, where he contributed to a range of mechatronic and control system solutions. He holds a Master of Science in Engineering Design with a specialization in Mechatronics from KTH Royal Institute of Technology, earned in 2016.
\end{IEEEbiography}


\begin{IEEEbiography}[{\raisebox{0pt}[1.2in][0pt]{\includegraphics[width=1in,height=1.25in,clip]{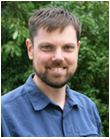}}}]{Oscar Flärdh} is a Product Owner at Traton Group R\&D. His current work focuses on parameter estimation and energy consumption functions. Prior to that, he has had various positions at Powertrain Control Systems Development at Scania CV AB in Södertalje, Sweden. That work contained various aspects of powertrain control algorithms, mainly focusing on energy efficiency.
He received the Ph.D. degree in Automatic Control from KTH Royal Institute of Technology, Stockholm, Sweden, in 2012 and the M.Sc. degree in electrical engineering and applied physics from Linköping University, Linköping, Sweden, in 2003.
\end{IEEEbiography}


\begin{IEEEbiography}[{\raisebox{0pt}[1.2in][0pt]{\includegraphics[width=1in,height=1.2in,clip]{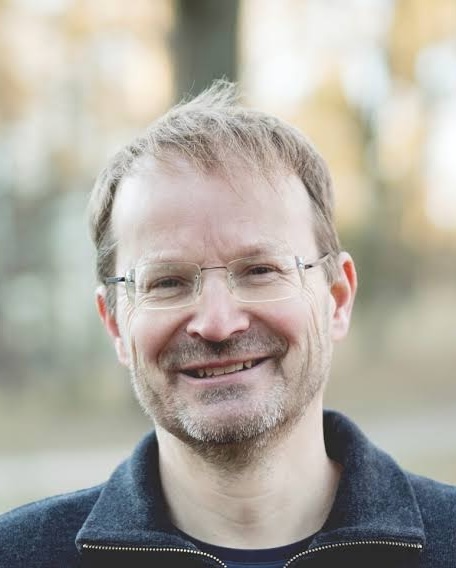}}}]{Håkan Hjalmarsson} was born in 1962. He received the M.S. degree in Electrical Engineering in 1988, and the Licentiate degree and the Ph.D. degree in Automatic Control in 1990 and 1993, respectively, all from Linköping University, Sweden. He has held visiting research positions at California Institute of Technology, Louvain University and at the University of Newcastle, Australia. He has served as an Associate Editor for Automatica (1996-2001), and IEEE Transactions on Automatic Control (2005-2007) and been Guest Editor for European Journal of Control and Control Engineering Practice. He is Professor at the Division of Decision and Control Systems, School of Electrical Engineering and Computer Science, KTH, Stockholm, Sweden and also affiliated with the Competence Centre for Advanced BioProduction by Continuous Processing, AdBIOPRO. He is an IEEE Fellow and past Chair of the IFAC Coordinating Committee CC1 Systems and Signals. In 2001 he received the KTH award for outstanding contribution to undergraduate education. He was General Chair for the IFAC Symposium on System Identification in 2018. His research interests include system identification, learning of dynamical systems for control, process modeling control and also estimation in communication networks.
\end{IEEEbiography}

\end{document}